%% file: main_arxiv.tex
\documentclass[a4paper,10pt]{article}
\usepackage[english]{babel}
\usepackage{amsmath}
\usepackage{siunitx}
\usepackage[a4paper,left=3cm,right=3cm,top=2.5cm,bottom=2.5cm]{geometry}
\usepackage{hyperref}
\hypersetup{
colorlinks=true,
linkcolor=blue,
urlcolor=red,
pdftitle={Valentin2ndPaper},
}
\usepackage{fancyvrb}
\usepackage{fancyhdr, lastpage}
\usepackage{bm}
\usepackage{pdfpages}
\usepackage{authblk}
\usepackage{graphicx}
\usepackage{float}
\usepackage{comment}
\usepackage{caption}
\usepackage{cite}
\usepackage[colorinlistoftodos]{todonotes}
\usepackage{mathtools}
\usepackage{amssymb}
\usepackage{amsthm}
\usepackage{multirow}
\usepackage{tabularx}
\usepackage{longtable}
\usepackage{setspace}
\usepackage{xcolor}
\usepackage{titlesec}
\usepackage{booktabs}
\usepackage[draft,inline,marginclue]{fixme}
\FXRegisterAuthor{gs}{ags}{\color{red}GS}

\DeclareGraphicsExtensions{.pdf,.png}
\usepackage[ruled, vlined, linesnumbered]{algorithm2e}
\usepackage{algpseudocode}
\usepackage{cleveref}
\usepackage{fixme}
\fxsetup{status=draft}
\graphicspath{ {./images/} }

\title{A hybrid reduced-order model for segregated fluid-structure interaction solvers in an ALE approach at high Reynolds number}
\author[1]{Valentin Nkana Ngan\footnote{vkanang@sissa.it}}
\author[2]{Giovanni Stabile\footnote{giovanni.stabile@santannapisa.it}}
\author[3]{Andrea Mola\footnote{andrea.mola@imtlucca.it}}
\author[1]{Gianluigi Rozza\footnote{grozza@sissa.it}}

\affil[1]{Mathematics Area, mathLab, SISSA, via Bonomea 265, I-34136 Trieste, Italy}
\affil[2]{The Biorobotics Institute, Sant'Anna School of Advacended Studies, V.le R. Piaggio 34, 56025, Pontedera, Pisa - Italy}
\affil[3]{MUSAM Continuum Mechanics Laboratory, Scuola IMT Alti Studi Lucca - Piazza S. Ponziano, 6 - 55100 Lucca, LU, Italy }
\date{\today} 

\begin{document}
\tableofcontents
\maketitle
\input{sections/abstract}

\listoffixmes
\input{sections/intro}
\input{sections/fsiformulation}
\input{sections/methodology}
\input{sections/resanddiscuss}
\input{sections/conclusion}
\input{sections/ackno}
\appendix
\input{sections/appendix}
\bibliographystyle{abbrv}
\bibliography{biblio}
\end{document}

%% file: sections/abstract.tex
\begin{abstract}
This study introduces a first step for constructing a hybrid reduced-order models (ROMs) for segregated fluid-structure interaction in an Arbitrary Lagrangian-Eulerian (ALE)
approach at a high Reynolds number using the Finite Volume Method (FVM). 
The ROM is driven by  proper orthogonal decomposition (POD) with hybrid techniques that combines the classical Galerkin projection and two data-driven methods (radial basis networks , and neural networks/ long short term memory). 
Results demonstrate the ROM’s ability to accurately capture the physics of fluid-structure interaction phenomena.  This approach is validated through a case study focusing on flow-induced vibration (FIV) of a pitch-plunge airfoil at a high Reynolds number ($Re=10^7$).
\end{abstract}

%% file: sections/intro.tex
\section{Motivation and state-of-the-art}
Interactions between fluid and moving boundaries are among the most essential issues of fluid dynamics. They arise in wind turbines, civil engineering (e.g., the influence of wind on bridges and buildings), and the aerospace industry \cite{stankiewicz2013arbitrary}. In the last case, fluid-structure interaction (FSI) plays an essential role in the design process of an aircraft. 
A significant amount of effort has been put into understanding and solving problems related to FSI using computational fluid dynamics (CFD).  
CFD has become an essential tool in the analysis and design of aerospace vehicles thanks to the advances in both computational algorithms and hardware. With these advances in CFD, the number of wings tested experimentally in the design of a typical commercial aircraft has decreased by an order of magnitude from the last four decades so an effective use of CFD is a key ingredient in the successful design of modern commercial aircraft \cite{Johnson2005}. 
Nowadays,  CFD tools can accurately predict aircraft aerodynamics in cruise conditions and complement wind tunnel and flight tests in the aircraft design process.
However, traditional high-fidelity aerodynamics simulations can be computationally too expensive for scenarios requiring real-time responses (e.g., flow control) and/or predictions for many different configurations (e.g., design-space exploration and flight-parameter sweep). 
The design of a new aircraft requires an analysis of a huge number of variants. One has to check different aircraft configurations, mass cases, gusts, and maneuvers giving (even with engineering experience for current configurations and technologies) hundreds of thousands of simulations \cite{rossow2008high}. 

Reduced-order models (ROMs) come into play to accelerate the solution of unsteady and/or parameterized aerodynamics problems in real-time and/or many-query scenarios \cite{2021}. The ROM approach in general and particularly based on the Proper Orthogonal Decomposition (POD) method has been extensively adopted for various engineering applications \cite{dao2021projection}. "ROMs are nowadays probably the most important mathematical technique for realizing a digital twin during the life cycle of a product. By using them, simulation models can be more interactive, reliable, continuous, accessible, and distributable" \cite{keiper2018reduced} i.e. ROM can be seen as a key enabler for a new generation of digital twins. "As a concrete example, one could consider the typical ROM problem: given a dynamical system, find a reduced dynamical system that approximates the original system with a controllable trade-off between error and speed, and preservation of key properties like stability."

The model-order reduction literature for aerodynamics problems is growing, with contributions from both engineering and applied mathematics communities.  The study of Anttonen et al.\cite{anttonen2001techniques, anttonen2005applications} outlined that an additional difficulty is reached when considering aeroelastic non-linear dynamical systems. In FSI cases, especially when adopting high-fidelity aerodynamics, some theoretical limits arise from the POD theory formulated in the deformable domain. The POD is based on a definition of a spatial correlation of the system. However, the snapshots resulting from an aeroelastic CFD solver cannot ensure this point as, notoriously, the mesh is moving and deforming during the simulation. The loss of the spatial correlation and the increasingly important stability and accuracy problems make the aeroelastic ROM study widely challenging. In their numerical examples involving a moving airfoil,  Freno et al. \cite{Freno2015} showed that when an index-based domain is used to build the ROM, similar to the one considered by Anttonen et al.\cite{anttonen2001techniques}, numerical simulations do not suffer from the mesh deformation limitation discussed above.
In the case of a rigid body motion as considered in this study, an interesting manner to avoid issues associated to mesh deformation are presented by Lewin et al. \cite{Lewin2005} and Placzek \cite{placzek2009construction}. They performed the projection of the governing equations in a non-inertial reference frame to preserve the POD formulation's consistency. However, in their case,  stability problems appear when considering highly non-linear flows.  Troshin et al. \cite{Troshin2016} outlined an alternative POD methodology for a flow field in a domain with moving boundaries. The moving domain is mapped to a stationary domain by combining a transfinite interpolation and an algorithm for volume adjustment. Liberge et al. \cite{liberge2010reduced} implemented a multi-phase method that allows the performance of the  POD on a moving domain using characteristic functions to follow the fluid-structure interface. Falaize et al. \cite{Falaize2019} extended such formulation for flows induced by rigid bodies in forced rotation. Also, they included parametric changes in the proposed model. Longatte et al. \cite{Longatte2018} explored the behavior of POD-multiphase ROM presented in  \cite{liberge2010reduced} when the parameter values are different from those used to build the POD basis. Stankiewicz et al. \cite{stankiewicz2008reduced, Stankiewicz2013} deepen the study of Anttonen et al. \cite{anttonen2001techniques} with test cases of increasing complexity also considering parametric changes. Freno et al. \cite{Freno2015, Freno2014} dealt with general non-linear systems in an aeroelastic context. They used dynamics basis functions to consider the domain deformation dynamics by defining dynamics related to the instantaneous deformed configuration for the projection basis. In addition, they considered a fully non-linear system, but since it is impossible to make it explicit, they used the FOM to evaluate the non-linear term at each time step. As a result, the computational efficiency of the ROM is reduced significantly.
Also, Shinde et al .\cite{shinde2019galerkin} extended the Galerkin-free approach to FSI problems by interpolating POD basis including mesh deformations. Alternatively, Thomas et al. \cite{Thomas2010} developed a nonlinear ROM for a dynamically nonlinear solver for limit cycle oscillation analysis. They used a nonlinear frequency domain harmonic balance method \cite{Hall2002} in conjunction with a Taylor series expansion and POD to create a frequency-domain nonlinear ROM.
In the context of system identification approaches, Chen et al. \cite{Chen2012} developed a support vector machine (SVM) based ROM for predicting the limit cycle oscillation induced by the nonlinear aerodynamics of an aeroelastic system. Mannarino et al.\cite{Mannarino2015, Mannarino2015rnns} developed a recurrent neural networks-based ROM technique in the discrete-time domain to deal with nonlinearities in FSI problems. Kou et al. \cite{Kou2017} derived a ROM for the investigation of limit cycle oscillations and flutter behaviors of an airfoil by combining linear auto-regressive with exogenous input (ARX) model with radial basis neural networks (RBFNNs) model for the nonlinear approximation. 
The current work is aligned in the same directions as the following studies \cite{miyanawala2019hybrid, gupta2022hybrid, miyanawala2018hybrid, dao2021projectionnns, mohan2018deep, whisenant2020galerkin}. This work differs from the works mentioned above by considering the motion of the mesh in the Arbitrary Lagrangian-Eulerian (ALE) sense. Moreover, it proposes an intrusive hybrid approach where the main partial differential equations (PDEs) are treated using a standard POD-Galerkin projection approach and radial basis functions networks for the grid motion interpolation.  More importantly, the current research expands upon our previous work done in  \cite{valnka}, by providing two data-driven approaches,
for predicting the eddy viscosity at the reduced-order level. 
This recipe results in a hybrid data-driven reduced-order model for FSI for any segregated solvers in the Finite Volumes Method (FVM). 
The proposed recipe combines the strengths of the POD-based reduced-order modelling and machine learning. In fact, by incorporating data-driven techniques, the ROM could  achieve high accuracy and efficiency, making it a powerful tool for simulating complex
FSI problems which could be useful in industrial applications in the development of digital twins systems. 
%

The structure of the manuscript is as follows:  The \cref{genmathform} begins with the formulation of the fluid-structure interaction with turbulence modeling. 
The section contains four subsections.  The first Sub\cref{structmotion} presents the structure motion. The following Sub\cref{fluidmotion}
deals with the mathematical formulation of the fluid's motion in the ALE setting, followed by the coupling strategy at the interface in Sub\cref{couplingcond}.  The \cref{NumDisFOMROM} presents the methodology carried out in this study. The section is divided into five subsections. The Sub\cref{standardfv} addresses the numerical discretization of the full-order model. The Sub\cref{rompro} discusses the reduced-order model concept followed by the discussion of POD for turbulence incompressible flows in  Sub\cref{redpimple} by insisting on its main properties in terms of model reduction while Sub\cref{rnns} discusses the machine learning algorithms for predicting the eddy viscosity.
In the last Sub\cref{pod-rbf}, the POD-RBF for interpolating the point cloud motion is introduced.
The \cref{resdiscuss} presents and discusses the numerical results obtained in this study.  Finally, a few considerations and possible avenues for future developments for this study are presented in \cref{conclusion}.

%% file: sections/fsiformulation.tex
\section{Governing equations of the fluid-structure interaction problem and turbulence modeling}\label{genmathform}
This section presents the mathematical formulation of the FSI model. The following assumptions are considered:
\begin{itemize}
    \item The fluid is viscous, incompressible and Newtonian; 
     \item The geometry and flow field considered are two-dimensional;
    \item The airfoil structure is rigid; 
    \item The elastic connection between the airfoil and the ground is represented by a set of linear and angular springs and dampers;  
    \item The airfoil undergoes a free translation (vertical direction) and rotational pitch motion.
\end{itemize}
Therefore, based on the  aforementioned assumptions, the  present part briefly describes the  high-dimensional full-order model to simulate the
coupled fluid-body interaction using the Navier-Stokes equations in the ALE setting, considering the rigid body dynamics and the coupling conditions at the interface.

\subsection{Structural dynamics equations}\label{structmotion}
In this work, the aeroelastic structural model as depicted in \cref{fig:airfoil} is governed by a two-degree freedom pitch and plunge system. 
The equations of the structure's motion are:
\begin{align}
	& m\ddot{h} + c_h\dot{h} +  k_h h - mb\ddot{\theta}\cos \theta + mb\dot{\theta}^2\sin \theta = F_h(t).\label{structmotion1} \\
	& \boldsymbol{I}_{\theta}\ddot{\theta} + c_{\theta}\dot{\theta} + k_{\theta}\theta - mb\ddot{h}\cos\theta  = \boldsymbol{M}_{\theta} (t).\label{structmotion2}
\end{align}
  $m$ being the mass of the airfoil per unit span,  $F_h(t)$ the sectional lift per unit span, $\boldsymbol{I}_{\theta}$ the sectional moment of inertia of the airfoil, $\boldsymbol{M}_{\theta}(t)$ is the pitching moment,  $\theta (t)$ the pitch rotation, h(t) the plunge displacement, $b$ the distance between the pivot location and the center of mass. The structural stiffness of the plunge and pitch is designated by $k_h$ and $k_{\theta}$; the related damping coefficients are $c_h$ and $c_{\theta}$.  The structural frequencies are $f_{\theta} = (2\pi)^{-1} \omega_{\theta}$, and $f_h = (2\pi)^{-1}\omega_h$  with $\omega_{\theta} = \sqrt{{k_{\theta}}/ {I_{zz}}}$ and $\omega_h = \sqrt{{k_h }/{m}}$. $I_{zz}$ being the moment of inertia $\boldsymbol{I}_{\theta}$ in the z-direction. For a thorough introduction to structural dynamics and aero-elasticity refer to \cite{hodges2011introduction}.
  
\subsection{Fluid dynamics equations and turbulence modeling}\label{fluidmotion}
The  unsteady Reynolds Average Navier-Stokes Equations  in the ALE framework for a \textit{Newtonian fluid} are written as follows:


\begin{align}
\nabla \cdot \boldsymbol{\bar{u}}  = 0.\label{eq1}\\
\frac{\delta \boldsymbol{\bar{u}} }{\delta t}+  \nabla \cdot (\boldsymbol{\bar{u}}\otimes(\boldsymbol{\bar{u}} -\boldsymbol{u}^g)) =  \frac{1}{\rho}\nabla \cdot \left[(\nu + \nu_t)\nabla \boldsymbol{\bar{u}})\right] -\frac{1}{\rho}\nabla \bar{p}.\label{eq2}\\
\frac{\delta \Omega (t)}{\delta t} +  \nabla \cdot \boldsymbol{u}^g = 0 \label{eq3}.
\end{align}
 $\boldsymbol{\bar{u}}$ is the average velocity field, $\bar{p}$ the average pressure field, and $\boldsymbol{u}^g$ the grid velocity.
 With $\nu$ being  the kinematic viscosity and  $\nu_t$ is the so-called turbulent viscosity. The quantity $\nu_t$ is suitably a viscosity only from the dimensional point of view and it is called viscosity considering the analogy of the Boussineq approximation and with the shear stress relations in a Newtonian fluid. The molecular viscosity is a property of the fluid and not its motion. 
A variety of methodologies are available in the literature to solve the eddy viscosity. This work uses the k-$\omega$ SST  (shear stress transport) introduced in \cite{menter2003ten}.
The time derivative in the ALE framework is given by:
\begin{align}
    \frac{\delta}{\delta t} = \frac{\partial}{\partial t} + \boldsymbol{u}^g\nabla.
\end{align}
The grid which moves in space must also obey the conservation law \cite{tsui2013finite}, which is stated as "the change in volume (area) of each control volume between time $t^n$ and $t^{n+1}$  must equal the volume (area) swept by the cell's boundary during $\Delta t = t^{n+1} -t^n$" which may be expressed as:
\begin{align}\label{gcl}
\frac{\delta }{\delta t}\int_{\Omega_i}d\Omega_i+  \int_{S_i}\boldsymbol{u}^g\cdot \boldsymbol{n} dS_i= 0 ~~~  \equiv ~~~ \displaystyle\frac{\delta \Omega_i}{\delta t} + \nabla \cdot \boldsymbol{u}^g = 0, 
\end{align}
for every control volume $\Omega_i = \Omega_i(t)$. $\boldsymbol{n}$ being the outward unit normal vector on the boundary surface, and $S_i = \partial \Omega_i$.
By multiplying \cref{gcl} by $\rho$ and using the incompressibility constraint leads to:  
\begin{align}
\int_{\partial \Omega_i} \boldsymbol{u}^g\cdot \boldsymbol{n} dS_i = 0.
\end{align}
This means there is no need to consider the grid velocity in the continuity equation. 
Additionally, there are \textit{initial and boundaries conditions}.  
This work uses URANS as it is the workhorse of turbulence modeling in industrial applications \cite{hijazi2020data}.

\subsection{Coupling conditions at the interface}\label{couplingcond}
The coupling between fluid and structure is achieved at the boundary conditions on the common interface $\Gamma (t)$ which all stem from simple physical principles: \textit{kinematic condition} (the fluid velocity, gird velocity,and structure's velocity are continuous at the interface),  \textit{dynamic condition} (the normal stresses of the fluid and structure are continuous on the interface), and the \textit{geometric condition} (the fluid and structure domain should always match.
\begin{align}
\boldsymbol{u}\cdot \boldsymbol{e}_y = \boldsymbol{u}_g\cdot \boldsymbol{e}_y  = \dot{h} ~~~ \text{and} ~~~\int_{\Gamma (t)}(\boldsymbol{\sigma}(\boldsymbol{x}, t)\cdot \boldsymbol{n})\cdot \boldsymbol{n}_y d\Gamma + F_h(t) = 0, 
\end{align}
with  $\boldsymbol{\sigma}(\boldsymbol{x}, t) = -p(\boldsymbol{x}, t)\mathbf{I} + \mu \left(\nabla \cdot \boldsymbol{u}(\boldsymbol{x}, t) + \left( \nabla \cdot \boldsymbol{u}(\boldsymbol{x}, t) \right)^T \right)$ as the fluid is assumed to be Newtonian. $F_h (t)$ is the time dependent lift force.

\subsection{Mesh motion strategy}\label{motionstrategies}
Translation and rotational motion of the centre of gravity (COG) are  accounted  by solving  Newton’s second law  \cref{structmotion1,structmotion2} in the global inertial reference frame.
After the linear and angular accelerations have been computed using \cref{structmotion1,structmotion2}, 
translation and rotational kinematics are used to update the body linear and angular velocities. After calculating the motion of the rigid body it is necessary to move the boundary as well as the  mesh surrounding the body in order to maintain a good quality mesh. In this work, the mesh deformation technique used is the so-called Slerp (Spherical Linear Interpolation)  as it is  better at handling translation and rotational (of a solid body in the three arises XYZ) mesh deformation when it comes to cell shearing \cite{jasak2007openfoam} and has great applications in computer vision.

%% file: sections/methodology.tex
\section{Numerical Methodology}\label{NumDisFOMROM}
The present section opens with a brief outline of the details of the standard FVM used for the FOM discretization. Relevant details of the reduced model are then presented, along with a description of the projection algorithm for the fully discrete equations. Finally, it describes Machine Learning algorithms (neural networks and recurrent neural networks) used for the online computation of the eddy viscosity, and of the radial basis interpolation combined with proper orthogonal decomposition used for the online computation of the grid nodal displacement field.

\subsection{Numerical discretization of the full-order model}\label{standardfv}
The standard FVM aims to discretize the system of partial differential equations written in integral form following \cite{moukalled2016finite}. The present uses a 2-dimensional tessellation. $N_{h}$ represents the dimension of the full-order model (FOM) which is the number of control volumes in the discretized problem. The following addresses the discretization methodology of the momentum and continuity equations. In particular, a segregated approach is used to solve the momentum and continuity equations inspired by  Rhie-Chow interpolation \cite{Rhie1983}. 

The referenced domain $\Omega (t) $ is divided into a tessellation $\mathcal{T}(t) = \{\Omega_i(t) \}_{i=1}^{N_h}$ so that every cell $\Omega_i(t)$ is a non-convex polygon 
and $\displaystyle\bigcup_{i=1}^{N_h} \Omega_i(t) = \Omega (t)$ and $\Omega_i(t) \cap \Omega_j(t) = \emptyset \quad \forall i\neq j$. In the following, to simplify the notation $\Omega_i = \Omega_i(t)$ and $S_i = \partial \Omega_i(t)$. $S_i$ being the total surface related to cell $\Omega_i$.
The unsteady-state momentum equation written in its integral form for every cell of the tessellation reads as follows:
\begin{align}
    \int_{\Omega_i}\frac{\delta \boldsymbol{\bar{u}}}{\delta t}d\Omega_i + \int_{\Omega_i}\nabla \cdot [\boldsymbol{\bar{u}}\otimes(\boldsymbol{\bar{u}} - \boldsymbol{u}^g)]d\Omega_i - \displaystyle\int_{\Omega_i} \nu_{eff}\nabla^2\boldsymbol{\bar{u}}d\Omega_i  + \int_{\Omega_i}\nabla \bar{p}d\Omega_i = 0.
\end{align}
$\nu_{eff} = \nu_l + \nu_t$ being the effective viscosity and the sum of the $\nu_l$  (molecular viscosity)  and turbulent viscosity. 
In the sequel, the full-order model is analyzed term by term.
\subsubsection{The pressure gradient term}
The pressure gradient term is discretized using Gauss's theorem.
\begin{align}\label{pressure}
     \int_{\Omega_i}\nabla \bar{p} d\Omega_i =  \displaystyle\int_{S_i}\bar{p} d\boldsymbol{S} \approx \displaystyle\sum_{j}\boldsymbol{S}_{ij}\bar{p}_{ij}, 
\end{align}
where $\boldsymbol{S}_{ij}$ is the oriented surface dividing the two neighbor cells $\Omega_i$ and $\Omega_j$ and $\bar{p}_{ij}$ is the pressure evaluated at the center of the face $\boldsymbol{S}_{ij}$.
\subsubsection{The convective term}
The convective term can be discretized as follows using Gauss's theorem.
\begin{align}
    \int_{\Omega_i}\nabla \cdot [\boldsymbol{\bar{u}}\otimes(\boldsymbol{\bar{u}} - \boldsymbol{u}^g)]d\Omega_i  & = \int_{\Omega_i}\nabla \cdot (\boldsymbol{\bar{u}}\otimes\boldsymbol{\bar{u}})d\Omega_i - \int_{\Omega_i}\nabla \cdot (\boldsymbol{\bar{u}}\otimes \boldsymbol{u}^g)d\Omega_i \\
    & = \int_{S_i}d\boldsymbol{S}\cdot (\boldsymbol{\bar{u}} \otimes \boldsymbol{\bar{u}}) -  \int_{S_i}d\boldsymbol{S}\cdot (\boldsymbol{\bar{u}} \otimes  \boldsymbol{u}^g)\\ 
    &   = \displaystyle\sum_{j \in S_i}\boldsymbol{\bar{u}}_{ij}\boldsymbol{F}_{ij} - \displaystyle\sum_{j \in S_i}\boldsymbol{\bar{u}}_{ij}({\boldsymbol{u}^g}_{ij} \cdot \boldsymbol{S}_{ij}).
\end{align}
Here,  $\boldsymbol{\bar{u}}_{ij}$ is the velocity evaluated at the center of the face $\boldsymbol{S}_{ij}$, and $\boldsymbol{F}_{ij} = \boldsymbol{\bar{u}}_{ij}\cdot \boldsymbol{S}_{ij}$ is the flux of the velocity through the face $\boldsymbol{S}_{ij}$.
This procedure underlines two considerations. 
The first one is that $\boldsymbol{\bar{u}}_{ij}$ is not straightly available in the sense that all the variables of the problem are evaluated at the center of the cells. At the same time, the velocity is evaluated at the center of the face. Many different techniques are available to obtain it. However, the basic idea behind them all is that the face value is obtained by interpolating the values at the center of the cells. The second clarification is about fluxes: during an iterative process for the resolution of the equations, they are calculated using the velocity obtained at the previous step so that the non-linearity is easily resolved.
\subsubsection{The diffusion term}
The diffusion term is discretized as follows:
\begin{align}\label{diffusion}
    \int_{\Omega_i} \nu_{eff} \nabla^2 \boldsymbol{\bar{u}} d\Omega_i  = {(\nu_{eff})}_i\int_{S_i}d\boldsymbol{S}\cdot  \left( \nabla \boldsymbol{\bar{u}} \right)d\Omega_i \approx \displaystyle\sum_{j} {(\nu_{eff})}_{ij} \boldsymbol{S}_{ij}\cdot (\nabla \boldsymbol{\bar{u}})_{ij}, 
\end{align}
where $(\nu_{eff})_i$ is the effective viscosity of the $i$-th cell, $(\nu_{eff})_{ij}$ is the effective viscosity evaluated at the center of the face $S_{ij}$, and $(\nabla \boldsymbol{\bar{u}})_{ij}$ is the gradient of $\boldsymbol{\bar{u}}_{ij}$ evaluated at the center of the face $ \bm{S}_{ij} $. 
As for the evaluation of the term $\bm{S}_{ij}\cdot (\nabla \boldsymbol{\bar{u}})_{ij}$ in \cref{diffusion}, its value depends on whether the mesh is orthogonal or non-orthogonal. Notice that the gradient of the velocity is not known at the face of the cell. The mesh is orthogonal if the line that connects two cell centers is orthogonal to the face that divides these two cells. 
For orthogonal meshes, the term  $\boldsymbol{S}_{ij}\cdot (\nabla \boldsymbol{\bar{u}})_{ij}$ is evaluated as follows:
\begin{align}\label{orthogonal-mesh}
 \boldsymbol{S}_{ij}\cdot (\nabla \boldsymbol{\bar{u}})_{ij} \approx \|\boldsymbol{S}_{ij}\|\frac{\boldsymbol{\bar{u}}_i - \boldsymbol{\bar{u}}_j}{\|\boldsymbol{d}_{ij} \|},
\end{align}
where $\bm{d}_{ij}$ represents the vector connecting the centers of cells of index $i$ and $j$. 
If the mesh is non-orthogonal, then a correction term has to be added to \cref{orthogonal-mesh}. 
In that case, one has to consider computing a non-orthogonal term to account for the non-orthogonality of the mesh as given by the following relation \cite{jasak1996error}:
\begin{align}\label{non-orthogonal-mesh}
\boldsymbol{S}_{ij}\cdot (\nabla \boldsymbol{\bar{u}})_{ij} = \|\boldsymbol{\pi}_{ij}\|\frac{\boldsymbol{\bar{u}}_i-\boldsymbol{\bar{u}}_j}{\|\boldsymbol{d}_{ij} \|} + \boldsymbol{k}_{ij} \cdot (\nabla \boldsymbol{\bar{u}})_{ij}.
\end{align}
Herein, $\boldsymbol{S}_{ij} = \boldsymbol{\pi}_{ij} + \boldsymbol{k}_{ij}$ and  $\boldsymbol{\pi}_{ij}$ is chose to be parallel to $\boldsymbol{S}_{ij}$ and $\boldsymbol{k}_{ij}$ to be orthogonal to $\boldsymbol{d}_{ij} $. The term $(\nabla \boldsymbol{\bar{u}})_{ij}$ is obtained through interpolation of the values of the gradient at the cell centers $(\nabla \boldsymbol{\bar{u}} _i)$ and $(\nabla \boldsymbol{\bar{u}}_j )$. 
The discretized forms of \cref{eq1,eq2,eq3} are written in a compact as follows:
\begin{align}\label{compactmatform}
    \left[ 
       \begin{array}{cc}
         \mathbf{A}_u   & \mathbf{B}_p  \\
           \nabla (\cdot) & \mathbf{0}
       \end{array}
       \right]
    \left[ 
       \begin{array}{c}
         \boldsymbol{\bar{u}}_h \\
           \boldsymbol{\bar{p}}_h
       \end{array}
\right]  =  \mathbf{0}.
\end{align}
The above system matrix has a saddle point structure which is usually difficult to solve using a coupled approach. For this reason, a segregated approach is used in this work where the momentum equation is solved with a tentative pressure and later corrected by exploiting the divergence-free constraint. Next, the main traits of PIMPLE algorithm is recalled.
\subsubsection{The PIMPLE algorithm}\label{pimplealgo}
The PIMPLE algorithm is a  mix of SIMPLE \cite{patankar1983calculation}, and  PISO \cite{issa1986solution} algorithms.  
This algorithm is mostly used for unsteady problems requiring a high Courant number or a dynamic mesh such as the one considered in this study.
To better understand the procedure of the  PIMPLE algorithm,  some crucial points about both algorithms are reported in the following, as they will be useful later during the online phase. 
Note that in this subsection,  the quantities $\boldsymbol{u}_h^{n*}$, and $\boldsymbol{p}_h^{n-1}$ are the  averaged terms coming from the Reynolds decomposition for velocity and pressure.

Starting with the SIMPLE algorithm the first step is to solve the discretized momentum equation considering the pressure field of the previous iterations. 
The momentum matrix is divided into diagonal and extra-diagonal parts so that the following holds:
\begin{equation}\label{MomemtumMatrix}
  \mathbf{A}_u   \boldsymbol{u}_h^{n*} = \mathbf{A}\boldsymbol{u}_h^{n*} -  \mathbf{H}(\boldsymbol{u}_h^{n*}),
\end{equation}
with $n$ being an index to identify a generic iteration and $ \mathbf{A}_u$ satisfying the following relation: 
\begin{equation}\label{MomentumPredictor}
\mathbf{A}_u   \boldsymbol{u}_h^{n*} = -\mathbf{B}_p\boldsymbol{p}_h^{n-1}. 
\end{equation}
By using \cref{compactmatform}, the momentum equation can be reshaped as follows:
\begin{align}\label{velstar}
   \mathbf{A}\boldsymbol{u}_h^{n*} = \mathbf{H}(\boldsymbol{u}_h^{n*}) - \mathbf{B}_p\boldsymbol{p}_h^{n*} \Rightarrow & \boldsymbol{u}_h^{n*} =  \mathbf{A}^{-1}\mathbf{H}(\boldsymbol{u}_h^{n*})- \mathbf{A}^{-1}\mathbf{B}_p\boldsymbol{p}_h^{n-1}.
\end{align}
In an iterative algorithm, the next step is introducing a small correction to the velocity and pressure field inside the \textit{inner loop}. 
Then, one can define the following relations:
\begin{align}\label{corrections}
  \boldsymbol{u}_h^n = \boldsymbol{u}_h^{n*} + \boldsymbol{u}'  ~~~~~~~~~~ \boldsymbol{p}_h^n = \boldsymbol{p}_h^{n-1} + \boldsymbol{p}'.
\end{align}
where $\boldsymbol{u}_h^{n*}$ does not satisfy the continuity equation, and $\boldsymbol{u}_h^{n}$ does. $\Box '$ are the corrections for both terms.
By inserting \cref{corrections} in \cref{velstar}, and rearranging terms give:
\begin{equation}\label{Mixterms}
   \boldsymbol{u}_h^n - \boldsymbol{u}' = \mathbf{A}^{-1}[\mathbf{H}(\boldsymbol{u}_h^{n}) - \mathbf{H}(\boldsymbol{u'})-\mathbf{B}_p\boldsymbol{p}_h^{n}+\mathbf{B}_p\boldsymbol{p}'] 
\end{equation}
From \cref{Mixterms}, one deduces a relation between $\boldsymbol{u}'$ and $\boldsymbol{p}'$:
\begin{equation}\label{uprime}
    \boldsymbol{u}' = \Tilde{\boldsymbol{u}}' - \mathbf{A}^{-1}\mathbf{B}_p\boldsymbol{p}', 
\end{equation}
with 
\begin{equation}\label{utildeprime}
    \Tilde{\boldsymbol{u}}' = \mathbf{A}^{-1}\mathbf{H}(\boldsymbol{u}').
\end{equation}
As the following relation holds thanks to \cref{MomentumPredictor} :
\begin{equation}
   \boldsymbol{u}_h^n = \mathbf{A}^{-1}[\mathbf{H}(\boldsymbol{u}_h^{n}) -\mathbf{B}_p\boldsymbol{p}_h^{n}] 
\end{equation}
With the use of \cref{uprime} and the divergence operator $\nabla (\cdot)$ applied to $\boldsymbol{u}_h^{n}$ in \cref{corrections} knowing $\boldsymbol{u}'$ from \cref{uprime},  one obtain an equation that directly relates $\boldsymbol{p}'$ and $\boldsymbol{u}_h^{n*}$:
\begin{equation}
    [\nabla (\cdot)]\left(\mathbf{A}^{-1}\mathbf{B}_p\boldsymbol{p}'\right) = [\nabla (\cdot)]\boldsymbol{u}_h^{n*} + [\nabla (\cdot)]\Tilde{\boldsymbol{u}}'.
\end{equation}
Which is basically the discretized Poisson equation for pressure (PPE) expressed in terms of the velocity and pressure corrections.
In the SIMPLE algorithm, the velocity corrections $\Tilde{\boldsymbol{u}}'$ are unknown as $\mathbf{H}(\boldsymbol{u}')$ are not too and hence neglected, implying the following relation:  
\begin{equation}\label{relvelstarandpprime}
    [\nabla (\cdot)]\left(\mathbf{A}^{-1}\mathbf{B}_p\boldsymbol{p}'\right) = [\nabla (\cdot)]\boldsymbol{u}_h^{n*}.
\end{equation}
Therefore, $\boldsymbol{p}'$ is expressed as the only function of $\boldsymbol{u}_h^{n*}$ in \cref{relvelstarandpprime}. Then the corrected pressure is entered again in \cref{velstar} in order to obtain a new velocity field $\boldsymbol{u}_h^{n*}$ and repeat the procedure until the pressure correction falls below a given \textit{tolerance} and the velocity satisfy both the continuity and momentum equation. 

As the $\Tilde{\boldsymbol{u}}'$ is neglected, the SIMPLE algorithm converges slowly and is used mainly for steady-state simulations. Furthermore, to avoid instabilities, relaxation factor $\alpha_p$ and $\alpha_u$ are introduced in the computation of $\boldsymbol{p}_h^{n}$ and $\boldsymbol{u}_h^{n*}$ as follows:
\begin{equation}
    \boldsymbol{p}_h^{n}  = \boldsymbol{p}_h^{n-1} + \alpha_p\boldsymbol{p}'.
\end{equation}
\begin{equation}
    \boldsymbol{u}_h^{n*} =  \mathbf{A}^{-1}\mathbf{H}(\boldsymbol{u}_h^{n*})- \alpha_u \mathbf{A}^{-1}\mathbf{B}_p\boldsymbol{p}_h^{n-1}.
\end{equation}

The PISO algorithm comes to play to speed up the convergence after neglecting $\Tilde{\boldsymbol{u}}'$ and \textit{computing the pressure correction}  $\boldsymbol{p}'$  using \cref{uprime}. $\boldsymbol{u}'$ is  computed  as follows:
\begin{equation}\label{1stpresscorr}
    \boldsymbol{u}' = - \mathbf{A}^{-1}\mathbf{B}_p\boldsymbol{p}'.
\end{equation}
Allowing the computation of $\Tilde{\boldsymbol{u}}'$ using \cref{utildeprime}. One defines a second velocity correction equation mirroring \cref{uprime} as follows:
\begin{align}\label{PISO}
    \boldsymbol{u}'' = \Tilde{\boldsymbol{u}}' - \mathbf{A}^{-1}\mathbf{B}_p\boldsymbol{p}''. 
\end{align}
As $\boldsymbol{u}''$ in \cref{PISO} satisfy the continuity equation, one define  also a second pressure correction equation as:
\begin{align}\label{2ndpresscorr}
      [\nabla (\cdot)]\left(\mathbf{A}^{-1}\mathbf{B}_p\boldsymbol{p}'' \right) =   [\nabla (\cdot)]\Tilde{\boldsymbol{u}}'.
\end{align}
To sum up, what the PISO algorithm does more than the SIMPLE algorithm is to add a second inner loop to correct pressure and velocity. This speeds up the convergence, allowing this algorithm to be used in a transient simulation.  Following the procedure described by \cref{1stpresscorr,PISO,2ndpresscorr} further corrections steps can be  added, increasing both the algorithm's convergence and computational cost. The essential steps of the  PIMPLE algorithm involving mesh motion is reported in \cref{alg:pimple}. In  \cref{alg:pimple},  the iterations within one time-steps are called \textit{outer iterations}, they are performed in an
\textit{outer loop} in which the coefficients and the source matrix of the discretized equations are updated. The operations performed on linear systems with fixed coefficients are
called instead \textit{inner iterations} and they occur in the so called \textit{inner loop}.

\begin{algorithm}
\caption{PIMPLE algorithm with dynamic mesh.}
\label{alg:pimple}
\SetKwInOut{Input}{Input}
\SetKwInOut{Output}{Output}
\Input{\text{Initial fields $\boldsymbol{u}_h^{n*}$, $\boldsymbol{p}_h^{n-1}$, $\nu_t^0$,  and $\boldsymbol{\delta}^0$ \Comment{$\boldsymbol{\delta}^0$ initial displacement}; }  } 
\Output{\text{$\boldsymbol{u}_h^{n}$, $\boldsymbol{p}_h^{n}$,  $\nu_t^n$,  and $\boldsymbol{\delta}^n$;}}
\While{ $t \leq t_{end}$ }
 {
      \While{$\text{No. outer corrections $\geq$ 2}  ~~ \text{and} ~~\text{Tol} \geq \text{maxTol} $}
      {
          Compute the forces; \Comment{Using $\boldsymbol{u}_h^{n*}$, $\boldsymbol{p}_h^{n-1}$}\;
          Solve the rigid body problem \cref{structmotion1,structmotion2}  \Comment{To obtain the new COG}\;
          Solve the mesh motion problem \Comment{To obtain $\boldsymbol{\delta}^n$ see \cref{motionstrategies}}\;
          $\mathbf{A}_u \boldsymbol{u}_h^{n*}$  \Comment{Assembling the momentum matrix \cref{MomemtumMatrix}}\;
          Solve $\mathbf{A}_u \boldsymbol{u}_h^{n*} = -\mathbf{B}_p\boldsymbol{p}_h^{n-1}$  \Comment{Momentum predictor \cref{MomentumPredictor} to obtain $\boldsymbol{u}_h^{n*}$}\;
          $ [\nabla (\cdot)]\left(\mathbf{A}^{-1}\mathbf{B}_p\boldsymbol{p}'\right) = [\nabla (\cdot)]\boldsymbol{u}_h^{n*}$ \Comment{Assembling the matrix of PPE \cref{relvelstarandpprime}}\;
          Solve $ [\nabla (\cdot)]\left(\mathbf{A}^{-1}\mathbf{B}_p\boldsymbol{p}'\right) = [\nabla (\cdot)]\boldsymbol{u}_h^{n*}$ \Comment{PPE to obtain $\boldsymbol{p}'$} \;
          $\boldsymbol{u}' \gets - \mathbf{A}^{-1}\mathbf{B}_p\boldsymbol{p}'$\Comment{ Momentum corrector \cref{1stpresscorr}} \;
          \While{$ \text{No. inner corrections }$ } 
          {   
              $[\nabla (\cdot)]\left(\mathbf{A}^{-1}\mathbf{B}_p\boldsymbol{p}''\right) = [\nabla (\cdot)]\Tilde{\boldsymbol{u}}'$  \Comment{Assembling the matrix for PPE  
                \cref{2ndpresscorr}}\;
              Solve $ [\nabla (\cdot)]\left(\mathbf{A}^{-1}\mathbf{B}_p\boldsymbol{p}''\right) = [\nabla (\cdot)]\Tilde{\boldsymbol{u}}'$ \Comment{Recursively to obtain $\boldsymbol{p}''$}\;
              $\boldsymbol{u}' \gets \Tilde{\boldsymbol{u}}' - \mathbf{A}^{-1}\mathbf{B}_p\boldsymbol{p}''$: \Comment{Momentum corrector \cref{PISO}} \;
              }
              {
             Solve turbulence and other transport quantities to obtain $\nu_t^n$\;
             Update tolerance\;
             }
              $\boldsymbol{u}_h^{n*} \gets \boldsymbol{u}'$ \;
              $\boldsymbol{p}_h^{n-1}\gets \boldsymbol{p}_h^{n-1} + \boldsymbol{p}'$ \;
       }
}
\end{algorithm}

\subsection{The reduced problem}\label{rompro}
The first subsection recalls the proper orthogonal decomposition for determining the modal basis functions, the next two subsections discuss Galerking projection and the machine learning algorithms to predict the eddy viscosity. The last subsection discusses the radial basis concept for predicting the mesh motion.
\subsubsection{The proper orthogonal decomposition}\label{POD}
The Proper Orthogonal Decomposition (POD) is used to construct the low-dimensional space. The POD is a compression technique where a set of numerical realizations (in time or parameter space) is reduced into a number of orthogonal basis (spatial modes) that capture the essential information suitably combined from previously acquired system data \cite{anttonen2005applications}.  In the sequel, the POD is formulated only in time space because in this  work the parameter dependency is implicit.

This work applies the POD to a group of realizations called snapshots. It consists of computing a certain number of full-order solutions $\boldsymbol{s}_i = \boldsymbol{s}(t_i)$ where $t_i \in \boldsymbol{T}$ for $i = 1, \cdots, N$. $\boldsymbol{T}$ being the training collection of a certain number $N$ of the time values, to obtain a maximum amount of information from this costly stage to be employed later on for a cheaper resolution of the problem. Those snapshots can be resumed at the end of the resolution all together into a matrix $\boldsymbol{S} \in \mathbb{R}^{N_h\times N}$. As already mentioned, $N_h$ is the number of control volumes in the discretized domain.
\begin{align}
    \boldsymbol{S} & = \left[\boldsymbol{s} (\boldsymbol{x},t_1), \dots, \boldsymbol{s} (\boldsymbol{x},t_{N}) \right].
\end{align}
The idea is to compute the ROM solution that can minimize the error denoted here by $E$ see \cref{Erom} between
the obtained realization of the problem and its high-fidelity counterpart. In the POD-Galerkin scheme, the reduced order solution is represented as follows:
\begin{align}\label{seqq}
\boldsymbol{s} (\boldsymbol{x},t) \approx  \boldsymbol{s}^{ROM} (\boldsymbol{x},t) = \displaystyle\sum_{i=1}^{N_r}a_i(t)\boldsymbol{\phi}_i(\boldsymbol{x}).
\end{align}
Where $N_r \ll N_h$ ($N_h$ is the number of cells in the computational domain) is a predefined number, namely the dimension of the reduced solution manifold,  $\boldsymbol{\phi}_i$ is a generic pre-calculated ortho-normal function depending only on the space while $a_i(t)$ is the temporal modal coefficients  satisfying the following conditions:  
\begin{align}
    & a_j(t)  =   \left(\boldsymbol{\phi}_j,  \boldsymbol{s} (\boldsymbol{x},t) \right)_{L^2 (\Omega)},
    & \boldsymbol{\phi}_j^T \bm{M} \boldsymbol{\phi}_i = \delta_{ij}.
\end{align}
$\bm{M}$ being  the mass matrix defined by the chosen inner product. In the case of $L_2$-norm and FVM $\bm{M}$ is a diagonal matrix containing the cell volumes. 
The best performing functions $\boldsymbol{\phi}_i$ in this case, are  the ones minimizing the  $L^2$-norm error $E$ between all the reduced-order solutions $\boldsymbol{s}^{ROM}_i$, $i = 1, \cdots, N$ and their high fidelity counterparts:
\begin{align}\label{Erom}
    E = \displaystyle\sum_{i=1}^{N}\|\boldsymbol{s}^{ROM}_i - \boldsymbol{s}_i\|_{L^2(\Omega)} =  \displaystyle\sum_{i=1}^{N}\|\boldsymbol{s}_i - \displaystyle\sum_{i=1}^{N_r}\left(\boldsymbol{s}_i, \boldsymbol{\phi}_i\right)_{L^2(\Omega)}\boldsymbol{\phi}_i \|_{L^2(\Omega (t_0))}.
\end{align}
It can be shown that solving a minimization problem based on \cref{Erom} is equivalent to solving the following eigenvalue problem \cite{Kunisch2002}:
\begin{align}\label{eigenproblem}
    \boldsymbol{C} \mathbf{V} = \mathbf{V}\boldsymbol{\lambda}.
\end{align}
 $\boldsymbol{C} \in \mathbb{R}^{N\times N}$ being the correlation matrix between all the different training solutions of the snapshot matrix $\boldsymbol{S}$, $\mathbf{V} \in \mathbb{R}^{N\times N}$ is the matrix whose columns are the eigenvectors, and $\boldsymbol{\lambda} \in \mathbb{R}^{N\times N}$  is a diagonal matrix whose diagonal entries are the eigenvalues.  The entries of the correlation matrix are defined as follows:
\begin{eqnarray}
\boldsymbol{C}_{ij} = \left(\boldsymbol{s}_i, \boldsymbol{s}_j \right)_{L^2\Omega (t_0))}.
\end{eqnarray}
$\Omega (t_0)$ being the reference configuration of the computational domain in the case of grid motion.
Note that the projection is performed with respect to $L_2(\Omega(t))$ while POD is computed with respect  to $L_2(\Omega(t_0))$. 
Using a POD strategy, the required basis functions are obtained through the resolution of the eigenproblem mentioned in \cref{eigenproblem}, obtained with the method of snapshots by solving \cref{Erom}. One can compute the required basis functions  as follows:
\begin{eqnarray}
    \boldsymbol{\phi}_i = \displaystyle\dfrac{1}{N\sqrt{\lambda_i}}\displaystyle\sum_{j=1}^{N}\boldsymbol{s}_jV_{ji} ~~~~~ \forall i = 1, \cdots, N.
\end{eqnarray}
All the basis functions are collected into a single matrix:
\begin{equation}
      \boldsymbol{\Phi} = \left[\boldsymbol{\phi}_1, \cdots, \boldsymbol{\phi}_{N_r} \right] \in \mathbb{R}^{N_h \times N_r}.
\end{equation}
Which is used to project the high fidelity problem onto the reduced subspace so that the final system dimension is $N_r$. 
This procedure leads to a problem requiring a computational cost that is much lower than the original problem. Next, the Galerkin projection, adapted in this study is presented.

\subsubsection{POD-Galerkin projection for velocity and pressure equations }\label{redpimple}
All the high-fidelity solutions are obtained by employing a segregated algorithm iterating the momentum and pressure equations until convergence is reached. The full-order model for both the velocity and pressure in the discretized form  is given as follows:
\begin{align}
\mathbf{A}_u \boldsymbol{u}_h = \boldsymbol{b}_u \label{discritizefullmodel1}, \\
\mathbf{B}_p \boldsymbol{p}_h = \boldsymbol{b}_p \label{discritizefullmodel2}.
\end{align}
With $\mathbf{A}_u \in \mathbb{R}^{dN_h\times dN_h}$, $\boldsymbol{u}_h \in \mathbb{R}^{dN_h}$, $\mathbf{B}_p \in \mathbb{R}^{N_h\times N_h}$ is the matrix operator for Poisson equation pressure (PPE) discussed in sub\cref{standardfv}, $\boldsymbol{p}_h \in \mathbb{R}^{N_h}$, and $d=2$ is the dimension of the computational domain. 
$N_h$ being the number of control volumes (cells) in the mesh, $\boldsymbol{b}_u$ and $\boldsymbol{b}_p$ are the respective source terms in the discretized form of the momentum equation and PPE. In this section, Galerkin projection (on the fully
discrete equations) is used for the construction of the reduced-order method.  Next, the reduced expansions of the velocity and pressure fields is introduced:
$\boldsymbol{u}_h (\boldsymbol{x},t) \approx \boldsymbol{u}_r (\boldsymbol{x},t)$ and $\boldsymbol{p}_h(\boldsymbol{x},t) \approx p_r(\boldsymbol{x},t)$, with
\begin{eqnarray}
\boldsymbol{u}_r (\boldsymbol{x},t) =  \displaystyle\sum_{i=1}^{N_u}a_i(t)\boldsymbol{\phi}_i(\boldsymbol{x}) = \boldsymbol{\Phi} \boldsymbol{a}^T\label{ueqq1},\\ 
p_r(\boldsymbol{x},t) = \displaystyle\sum_{i=1}^{N_p}b_i(t)\boldsymbol{\xi}_i(\boldsymbol{x}) = \boldsymbol{\Xi} \boldsymbol{b}^T \label{ueqq2}.
\end{eqnarray}
Herein,  $a_i(t)$ and  $b_i(t)$ are modal coefficients;  $\boldsymbol{\phi}_i$ and $\boldsymbol{\xi}_i$ are the basis functions corresponding to the  POD modes of the velocity and pressure fields stored respectively in $\boldsymbol{\Phi} \in \mathbb{R}^{dN_h\times N_u}$ and $\boldsymbol{\Xi} \in \mathbb{R}^{N_h\times N_p}$ with $N_u$ and $N_p$ being the numbers of basis functions selected for the predicted velocity and pressure solutions respectively, $\boldsymbol{a} \in \mathbb{R}^{N_u}$ is the vector containing the coefficients for the velocity expansion while the same reads for pressure with respect to $\boldsymbol{b} \in \mathbb{R}^{N_p}$. 
For the construction of the reduced basis spaces, the POD strategy mentioned in sub\cref{rompro} is used on the snapshot matrices of the velocity and pressure fields to obtain two separate families of reduced basis functions.
\begin{align}
& \boldsymbol{\Phi} = \left[\boldsymbol{\phi}_1,\dots, \boldsymbol{\phi}_{N_u} \right] \in \mathbb{R}^{dN_h\times N_u}, \label{basismatrices1}\\
& \boldsymbol{\Xi} = \left[\boldsymbol{\xi}_1, \dots, \boldsymbol{\xi}_{N_p} \right] \in \mathbb{R}^{N_h\times N_p}\label{basismatrices2}.
\end{align}
The  \cref{discritizefullmodel1,discritizefullmodel2} are projected using in \cref{basismatrices1,basismatrices2} respectively leading to: 
\begin{align}
 \boldsymbol{A}_u^r\boldsymbol{a} = \boldsymbol{b}_u^r, \label{redsystems1}\\
 \boldsymbol{A}_p^r\boldsymbol{b} = \boldsymbol{b}_p^r\label{redsystems2}.
\end{align}
Where $\boldsymbol{A}_u^r = \boldsymbol{\Phi}^T\mathbf{A}_u\boldsymbol{\Phi} \in \mathbb{R}^{N_r^u\times N_r^u}$,    $\boldsymbol{A}_p^r = \boldsymbol{\Xi}^T\mathbf{A}_p\boldsymbol{\Xi} \in \mathbb{R}^{N_r^p\times N_r^p}$, $\boldsymbol{b}_u^r = \boldsymbol{\Phi}^T\boldsymbol{b}_u \in \mathbb{R}^{N_r^u}$, and $\boldsymbol{b}_p^r = \boldsymbol{\Xi}^T\boldsymbol{b}_p \in \mathbb{R}^{N_r^p}$.
The resulting \cref{redsystems1,redsystems2} can be solved using any method for dense matrices. In this work,  the Householder rank-revealing QR decomposition of a matrix with full pivoting is used and it is available in the Eigen library \cite{eigenweb}. 
As the main idea here is to rely on a method capable of being as coherent as possible concerning the high-fidelity problem  (\cref{alg:pimple}), in the following the  main steps for the reduced algorithm related to incompressible turbulent flows with mesh motion are reported in \cref{alg:redpimple}.

\begin{algorithm}
\caption{Reduced-PIMPLE algorithm with dynamic mesh.}
\label{alg:redpimple}
\SetKwInOut{Input}{Input}
\SetKwInOut{Output}{Output}
\Input{\text{Initial fields $\boldsymbol{u}_h^{n*}$, $\boldsymbol{p}_h^{n-1}$, $\nu_t^0$,  and $\boldsymbol{\delta}^0$  \Comment{$\boldsymbol{\delta}^0$ is the initial node displacement}; } } 
\Output{\text{ $\boldsymbol{u}_h^{n}$, $\boldsymbol{p}_h^{n}$, $\nu_t^n$,  and $\boldsymbol{\delta}^n$;}}
\While{ $t \leq t_{end}$ }
 {
      \While{$\text{No. outer corrections $\geq$ 2}  ~~ \text{and} ~~\text{Tol} \geq \text{maxTol} $}
      {
          Compute the forces; \Comment{Using $\boldsymbol{u}_h^{n*}$, $\boldsymbol{p}_h^{n-1}$}\;
          Solve the rigid body problem \cref{structmotion1,structmotion2}  \Comment{To obtain the new COG $\boldsymbol{\vartheta}_{new}$}\;
          Compute $\boldsymbol{c} = RBF (\boldsymbol{\vartheta}_{new})$ \cref{ThetaNew}\;    
          Reconstruct $\boldsymbol{\delta}^n  = \Psi \boldsymbol{c}^T$ \cref{pod-rbfs}\;
          $\mathbf{A}_u \boldsymbol{u}_h^{n*}$ \Comment{Assembling the momentum matrix \cref{MomemtumMatrix}}\;
          Solve $\boldsymbol{\Phi}^T \mathbf{A}_u \boldsymbol{\Phi} \boldsymbol{a}^{*} =\boldsymbol{\Phi}^T \mathbf{b}_u$ \Comment{To obtain $\boldsymbol{a}^{*}$ \text{with} $\mathbf{b}_u =-\mathbf{B}_p\boldsymbol{p}_h^{n-1}$} \;
          Reconstruct $\boldsymbol{u}_h^{n*}$\  \Comment{Using $\boldsymbol{a}^{*}$}\;
          $ [\nabla (\cdot)]\left(\mathbf{A}^{-1}\mathbf{B}_p\boldsymbol{p}'\right) = [\nabla (\cdot)]\boldsymbol{u}_h^{n*}$\Comment{Assembling the matrix of PPE \cref{relvelstarandpprime}}\;
          Solve $\boldsymbol{\Xi}^T \mathbf{A}_p \boldsymbol{\Xi} \boldsymbol{b}' = \boldsymbol{\Xi}^T\boldsymbol{b}_p$ \Comment{To obtain $\boldsymbol{b}'$}\;
          Reconstruct $\boldsymbol{p}'$\  \Comment{Using $\boldsymbol{b}'$}\;
          $\boldsymbol{u}' \gets - \mathbf{A}^{-1}\mathbf{B}_p\boldsymbol{p}'$ \Comment{Momentum corrector \cref{1stpresscorr}} \;
          \While{\text{No. inner corrections } } 
          {   
              $[\nabla (\cdot)]\left(\mathbf{A}^{-1}\mathbf{B}_p\boldsymbol{p}''\right) = [\nabla (\cdot)]\Tilde{\boldsymbol{u}}'$  \Comment{Assembling the matrix for PPE  
                \cref{2ndpresscorr}}\;
              Solve $\boldsymbol{\Xi}^T \mathbf{A}_p \boldsymbol{\Xi} \boldsymbol{b}'' = \boldsymbol{\Xi}^T\boldsymbol{b}_p$ \  \Comment{Recursively to obtain $\boldsymbol{b}''$ \text{where} $\boldsymbol{b}_p = [\nabla (\cdot)]\Tilde{\boldsymbol{u}}'$}\;
              Reconstruct $\boldsymbol{p}''$\  \Comment{Using $\boldsymbol{b}''$}\;
              $\boldsymbol{u}' \gets \Tilde{\boldsymbol{u}}' - \mathbf{A}^{-1}\mathbf{B}_p\boldsymbol{p}''$: \Comment{Momentum corrector \cref{PISO}} \;
           }
           {
            Evaluate the networks using NNs or LSTM \;  
            Reconstruct the new turbulent viscosity $\nu_t^n$\;
           }
           $\boldsymbol{u}_h^{n*} \gets \boldsymbol{u}'$ \;
           $\boldsymbol{p}_h^{n-1}\gets \boldsymbol{p}_h^{n-1} + \boldsymbol{p}'$ \;
       }
}
\end{algorithm}

\subsubsection{Machine learning for  eddy viscosity prediction}\label{rnns}
In this work, a deep neural network is used for the prediction of the eddy viscosity mimicking the works done in \cite{hijazi2020data, zancanaro2021hybrid, dao2021projectionnns}. In \cite{hijazi2020data} radial basis functions were used to predict the temporal coefficients of the  eddy viscosity based on the temporal coefficients of the velocity. The study carried out in \cite{zancanaro2021hybrid} used a fully connected neural network to predict the parameterized coefficients of the eddy viscosity in a steady simulation. Scholars in \cite{dao2021projectionnns} studied  the effects of the effective viscosity
in projection-based ROM simulations and employed a simple spline interpolation for the prediction of the first dominant mode of the temporal coefficient of the eddy viscosity from known values.

A fully connected neural network and a recurrent neural network based on the LSTM are presented here to predict temporal coefficients of eddy viscosity. The choice of  the aforementioned methodologies relies on the idea of building a reduced problem independent of the turbulent technique ($k-\epsilon$, $k-\omega$, etc.) that might be used in the original problem to calculate the eddy viscosity.
The first method uses the physical relation between the velocity field and eddy viscosity thanks to the Boussinesq hypothesis. The second method assume a one-to-one dynamical mapping between the low-dimensional states $\boldsymbol{n}^{i}$ and $\boldsymbol{n}^{i-1}$ defined in \cref{NutPOD}.
The low-dimensional eddy viscosity is approximated first using the POD decomposition on the snapshot matrix $\boldsymbol{S}_{\nu_t} \in \mathbb{R}^{N_h \times N_s}$ as :
\begin{align}\label{NutPOD}
\nu_t (\boldsymbol{x}, t)  \approx \displaystyle\sum_{i = 0}^{N_{\nu_t}}n_i (t)\boldsymbol{\psi}_i(\boldsymbol{x}) = \boldsymbol{\Psi} \boldsymbol{n}^T.
\end{align}
$\psi_i(\boldsymbol{x})$ and $n_i (t)$ are the POD spatial and temporal coefficients modes for eddy viscosity respectively.  $N_{\nu_t} \ll N_s$ denotes the selected number of modes to predict the eddy viscosity.
In contrast to the temporal coefficients modes of the velocity and pressure obtained by projecting the FOM onto the respective POD spatial modes and subsequently solving the reduced problem, the predicted temporal coefficients modes for the eddy viscosity are modeled via a multi-layer feed-forward neural network or a recurrent network. 


\begin{itemize}
    \item \textbf{Neural networks: }
 Within this method,  the neural networks are fed with temporal coefficients of the velocity field $\boldsymbol{a}$ and mapped to the temporal coefficient of the turbulent viscosity $\boldsymbol{ \Tilde{n}}$.  $\boldsymbol{ \Tilde{n}}$ being the prediction from the neural network. For a comprehensive description, the reader could refer to Goodfellow et al. \cite{Heaton2017}. 
In a feed-forward neural network, the output of a single layer of a given input $\mathbf{A} \in \mathbb{R}^N$ given by $\mathbf{Y} = f (\mathbf{W}\mathbf{A} + \mathbf{b}).$  $\mathbf{W} \in \mathbb{R}^{M\times N}$ being the weight matrix, $\mathbf{b} \in \mathbb{R}^M$ is a bias term and f (.) is a \textit{nonlinear function} that acts element-wise on its inputs. The Multi-layer neural networks are generated by feeding the output $\mathbf{h}_l = f_{l+1} (\mathbf{W}_l\mathbf{A}_l + \mathbf{b}_l)$ of a layer $l$ as the input of the next layer.  
The vector  $\mathbf{h}_l$ is often referred to as the hidden state or feature vector at the l-th layer. Generally, training a network
involves finding the parameters  $\boldsymbol{\theta} = \{\mathbf{W}_l,  \mathbf{b}_l\}_{l=0}^{l=L-1}$ such that the expected loss $\mathcal{L}(\Hat{\mathbf{Y}}, \mathbf{Y})$ between the output $\mathbf{Y}$ and the target value $\Hat{\mathbf{Y}}$ is minimized i.e.
\begin{align}
    \boldsymbol{\theta}_{*} & =  \arg \displaystyle\min_{\boldsymbol{\theta}}[\mathcal{L}(\Hat{\mathbf{Y}}, \mathbf{Y})],
\end{align}
where $\mathcal{L}(\Hat{\mathbf{Y}}, \mathbf{Y})$ is some measure of discrepancy between the predicted and target outputs given in some cases by:
\begin{align}
    \mathcal{L}(\Hat{\mathbf{Y}}, \mathbf{Y}) = \left[\frac{1}{n-1}\displaystyle\sum_{i=1}^{n}\frac{||\boldsymbol{Y}_i -  \boldsymbol{g}(\boldsymbol{Y}_{i-1}; \boldsymbol{\theta})  ||_2^2}{||\boldsymbol{Y}_i ||_2^2}   \right].
\end{align}
For example in the case of recurrent neural networks with $\boldsymbol{g}$ a
one-to-one mapping giving by:  $\boldsymbol{\Hat{Y}}_{i} = \boldsymbol{g}(\boldsymbol{Y}_{i-1}; \boldsymbol{\theta})$.


\item \textbf{Recurrent neural networks: }
This method assumes the following relationships:
\begin{align}
     \nu_t^{i+1} = \boldsymbol{g}(  \nu_t^{i}) &  \Rightarrow  \Tilde{\boldsymbol{n}}^{i+1} = \boldsymbol{g}(\Tilde{\boldsymbol{n}}^{i} ).
\end{align}
$\boldsymbol{g}$ being an unknown mapping between the flow fields of adjacent time steps. The goal consists of learning a dynamical operator $\boldsymbol{g}$ by finding the parameters $\boldsymbol{\theta}$ (weights and biases) that allow to recurrently predict finite time series of the low-dimensional states. Recurrent neural networks (RNNs) are designed specifically to deal with sequential data. Data coming from transient CFD simulations or any nonlinear dynamical system is a set of time series that is sequential. This makes the application of RNNs for non-linear dynamical systems quite relevant \cite{bukka2020data}.
RNNs differ from the traditional feed-forward neural network with the presence of a
recurrent connection. This particular recurrent connection is responsible for storing the history of previous inputs ($\boldsymbol{X}_{i}$) via hidden states. The basic mathematical structure of simple RNNs cell for an input $\mathbf{A}_i \in \mathbb{R}^N$ and output $\mathbf{Y}_i \in \mathbb{R}^M$ given as follows:
\begin{align} 
 \mathbf{X}_i = f (\mathbf{W}_x\mathbf{X}_{i-1} + \mathbf{W}_a\mathbf{A}_{i} +  \mathbf{b})  ~~~ \text{and} ~~~   \mathbf{Y}_i = g(\mathbf{W}_y\mathbf{X}_{i}).
\end{align}
$\mathbf{W}_x \in \mathbb{R}^{N_x\times N_x}$, $ \mathbf{W}_a \in N_x\times N$,  and $\mathbf{W}_y \in \mathbb{R}^{N_x\times M}$ being the hidden, input and
output weight matrices respectively.  $\mathbf{b} \in \mathbb{R}^M$ represents the bias term, and $\boldsymbol{X}_{i-1}$ the cell state at time $i-1$. RNNs are typically
trained using stochastic gradient descent (SGD), or some variant, but the gradients are calculated using the backpropagation through time (BPTT) algorithm \cite{werbos1990backpropagation}. For a detailed discussion on BPTT the reader might refer to \cite{bengio1994learning}.
This work considers RNNs equipped with LSTM (long short-term memory) units \cite{sak2014long}. The straightforward  prediction of  $\Tilde{\nu}_t$ at the online level is given using  \cref{NutPOD} i.e.
\begin{align}\label{NutPODNut}
    \Tilde{\nu}_t = \Psi \boldsymbol{\Tilde{n}}^T.
\end{align}

\end{itemize}

\subsubsection{POD with interpolation for mesh motion prediction}\label{pod-rbf}
This section presents a method to reduce the computational cost associated with the mesh motion part in the system. The advantage of this methodology is to make the online part independent of the mesh motion technique used at the offline stage. The methodology combines proper orthogonal decomposition with radial basis functions (RBF) networks on the point displacement field.  
The interpolation using RBF is given  by the following formula: 
\begin{equation}\label{rbfnns}
    f(\boldsymbol{x}_j)  = \displaystyle\sum_{j=1}^{N}\boldsymbol{w}_j\rho(|| \boldsymbol{x}_j - \boldsymbol{x}_k ||), 
\end{equation}
where  $f: \mathbb{R} \mapsto \mathbb{R}$ a known map at some finite number of points $f(\boldsymbol{x}_k) = y_k$, $k = 1, \cdots, N$, $\rho$ is a radial basis function,  $N$ is the number of neurons in the hidden layer, and $\boldsymbol{w}_j$ being the weight of neuron $j$ in the linear output neuron.  \cref{rbfnns} can be rewritten as a linear system, namely:
 \begin{equation}\label{rbfweights}
   \boldsymbol{y}   = \mathbf{G} \boldsymbol{w}^T, 
 \end{equation}
 where $ \mathbf{G} = (g_{kj}) = \rho(||\boldsymbol{x}_k - \boldsymbol{x}_j ||)$ being the Gram matrix. The weights are given by: $\boldsymbol{w} = \mathbf{G}^{-1}\boldsymbol{y}$.
The rationale behind POD-RBF regards the evaluation of the new temporal or parameterized coefficients 
computed at the parameter points $\boldsymbol{\vartheta}_k$; in this work, it will be a two-dimensional vector of the pitch and plunge  at a given time t. 
Next, the separability's assumption on the point displacement field is given as follows:
\begin{equation}\label{POD-displacement}
    \boldsymbol{\delta} ( \boldsymbol{x}, t_k )  \approx \displaystyle\sum_{i=1}^{N_{\delta}}c_i(\boldsymbol{\vartheta} (t_k))\boldsymbol{\psi}_i(\boldsymbol{x}), ~~~  \forall \boldsymbol{\vartheta} (t_k)=\vartheta_k \in \boldsymbol{\Theta}.
\end{equation}
$\boldsymbol{\Theta}$ being the training set of the airfoil motion at the offline stage, and $N_{\delta}$ the require number of basis to approximate the $\boldsymbol{\delta}$. 
In the online stage, as input, a new time-parameter value $\boldsymbol{\vartheta}_{new}$ is given,  and using the trained weights obtained from \cref{rbfweights},  the coefficient $c_i =  c_i(\vartheta_{new})$ is obtained by interpolation with RBF and $\boldsymbol{\vartheta}_{new}$ is obtained by solving  \cref{structmotion1,structmotion2} at the online stage. Then, the new coefficient is computed by:
\begin{equation}\label{ThetaNew}
    c (\boldsymbol{\vartheta}_{new}) =  \displaystyle\sum_{j=1}^{N_{\delta}}\boldsymbol{w}_j\rho(||\boldsymbol{\vartheta}_{new} - \boldsymbol{\vartheta}_j ||).
\end{equation}
The new point displacement is predicted as follows:
\begin{equation}\label{pod-rbfs}
    \boldsymbol{\delta} (\boldsymbol{x}, t_{new})  = \displaystyle\sum_{i=1}^{N_{\delta}}c_i(\boldsymbol{\vartheta} )\boldsymbol{\psi}_i(\boldsymbol{x}) =  \Psi \boldsymbol{c}^T.
\end{equation}

%% file: sections/resanddiscuss.tex
\section{Definition of test case,  simulation results, and discussion}\label{resdiscuss}
This section shows the results obtained for our reference test case which represents two-dimensional turbulent flow past a plunging and pitching airfoil. The simulation is carried out for a total time of 500 flow through times (FTTs). In the present case, such time unit is defined as the time $\text{FTT} = \dfrac{L}{||\boldsymbol{U}_{\infty}||} = 0.01$ required by a fluid particle to travel a distance equivalent to the airfoil chord $L$ at the speed of the undisturbed stream $||\boldsymbol{U}_{\infty}||$ \cite{komen2014quasi}.

\begin{figure}[H]
\centering
	\begin{tabular}{cc}
\multicolumn{2}{c}{\includegraphics[width=0.5\textwidth]{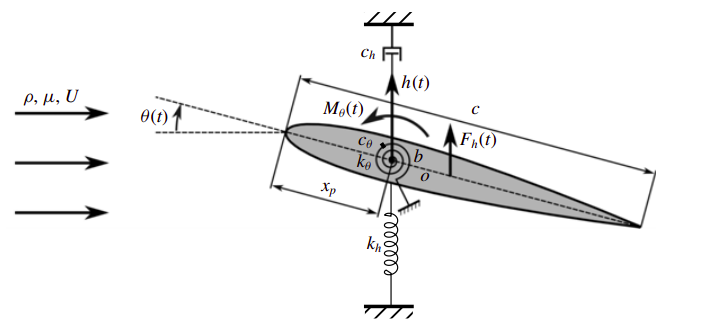}}\\
\multicolumn{2}{c}{(a)} \\
\includegraphics[width=0.4\textwidth]{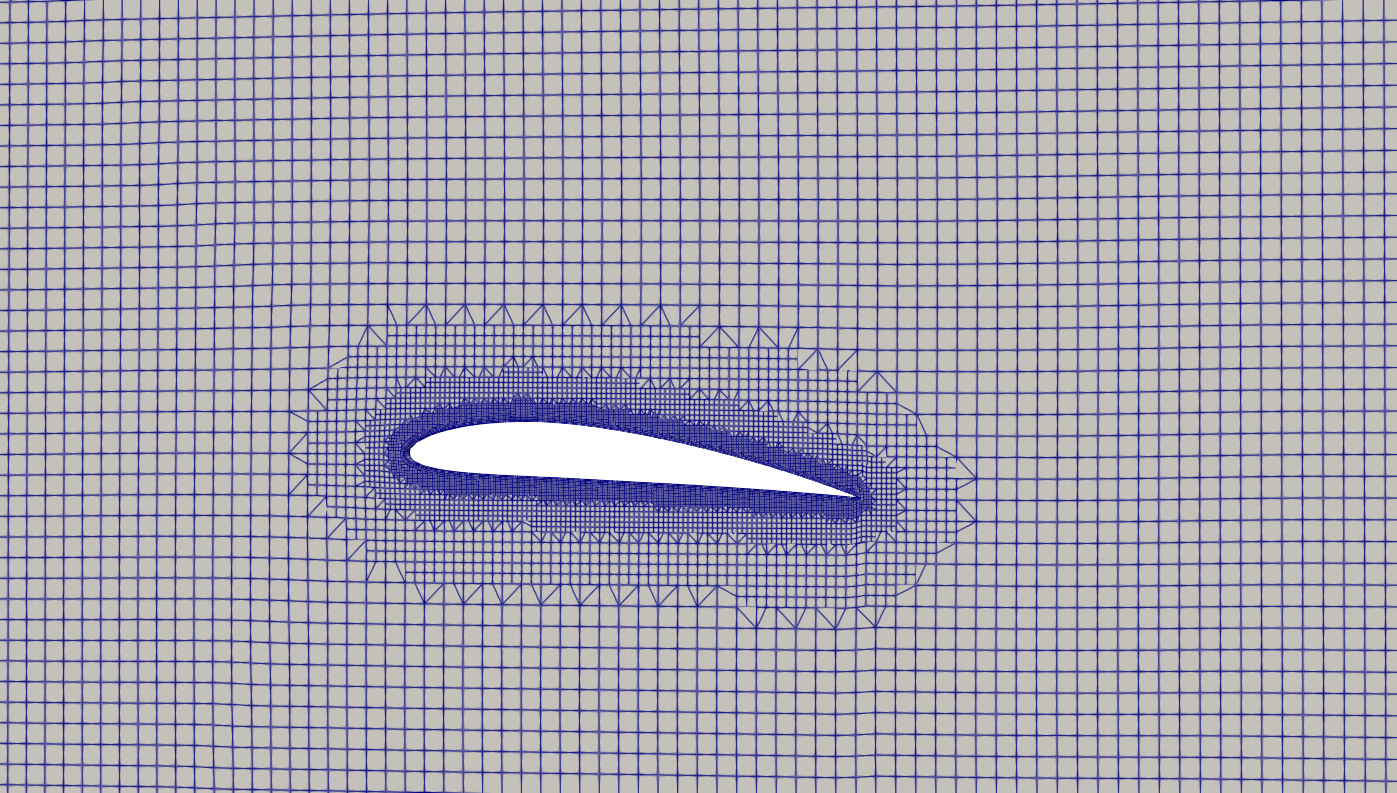}
        &
\includegraphics[width=0.33\textwidth]{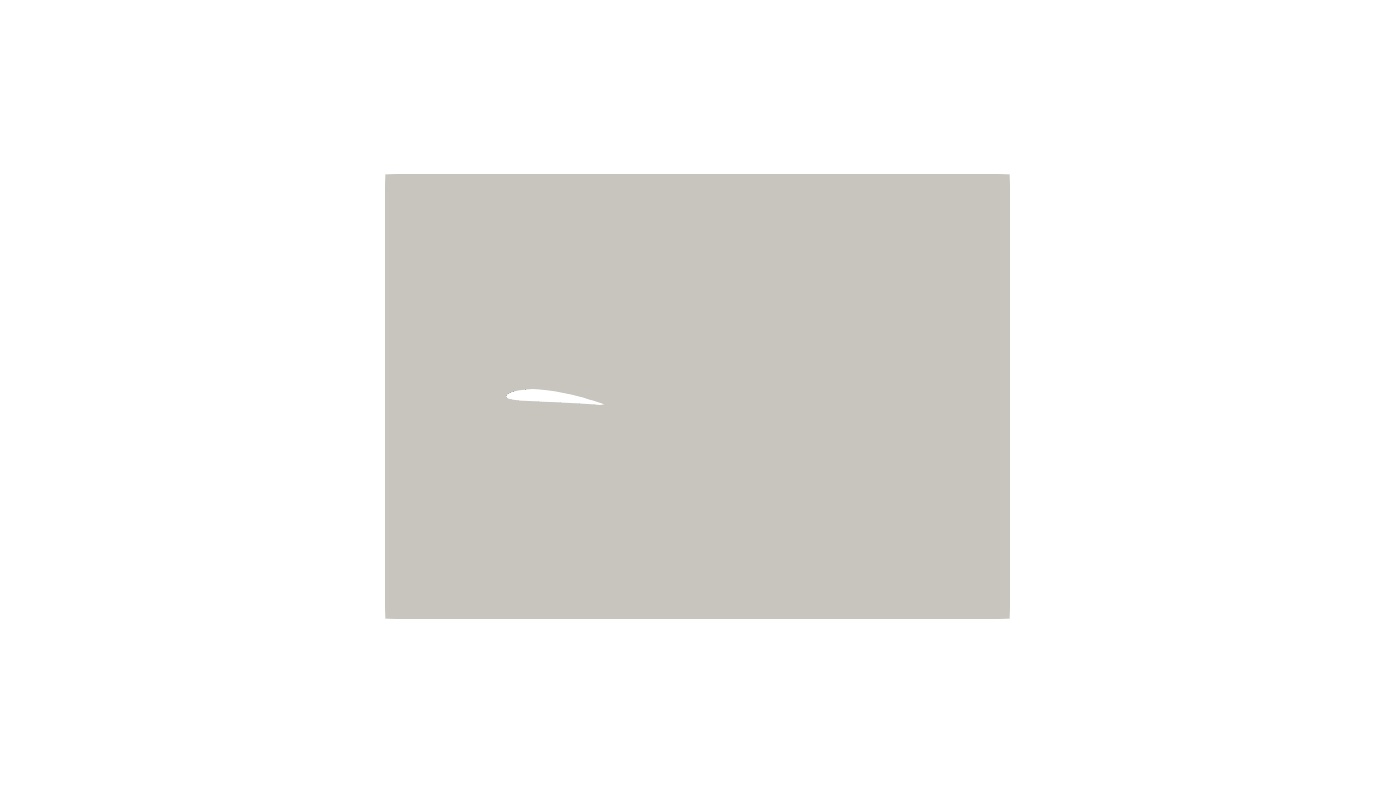}
\\                                                     
(b) & (c)
	\end{tabular}
	\caption{(a) Schematic of the fluid-structure system considered: a foil allowed to undergo 2 degrees of freedom fully passive plunging and pitching motion with spring constraints \cite{Wang2020}, (b) a picture of the zoomed mesh with 12 556 cells
(control volumes) and 26 316 node points 
near an  airfoil of chord length \SI{1.0}{\metre},  (c) picture showing the position of the foil in the computational domain}	
  \label{fig:airfoil}
\end{figure}

The 500 FTTs duration choice allows for the flow to fully develop also in the wake region. The test case of the study is the analysis of a two-degree freedom flutter as shown in \cref{fig:airfoil}.

\subsection{Definition of the test case}
Fig. \ref{fig:airfoil} (c) shows the 2D computational domain used in this work. The grid features 12 556 cells (control volumes) and 26 316 node points. It also features an extrusion layer around the airfoil to better capture the physical boundary layer. The molecular viscosity $\nu = \SI{e-5}{\metre^2\per{\s}} \approx \nu_{air}$.


The boundary conditions prescribed for the velocity field at the inflow boundary are non-homogeneous Dirichlet. The velocity value imposed is that of a uniform and constant horizontal velocity $\boldsymbol{U}_{\infty} = (U_{in}, 0)$ with $U_{in} = \SI{e+2}{\metre\per \s}$. Given the airfoil's chord length  $L = \SI{1.0}{\metre}$, the resulting Reynolds number is $10^7$. On the --- moving --- airfoil boundary a Dirichlet boundary condition is applied, imposing that the fluid's velocity is equal to the airfoil surface one. On the top and bottom boundaries, we made use of symmetry boundary conditions, while, zero pressure value and zero normal velocity gradient are prescribed at the outflow  boundary.
The full-order simulations are carried out using the PIMPLE algorithm, as described in \cref{standardfv}.
The PIMPLE algorithm can adapt the time step in a way that assures the maximum Courant number does not exceed a prescribed value which in this case, has been set to  $0.5$. As Reynolds average Navier-Stokes (RANS) is used in this work, the time step ${\Delta t}$ is chosen based on the following \textit{rule of thumb} \cite{komen2014quasi}: ${\Delta t} \approx 100{\Delta t}_{DNS}$.
\begin{eqnarray}\label{CoNum}
    \text{with} ~~ {\Delta t}_{DNS} \approx \frac{Co \times \eta}{U_{in} } ~~ \text{and} ~~ \eta  \equiv L\times Re^{-\frac{3}{4}}, 
\end{eqnarray}
where $L$ is the size of the largest eddies (in this work, $L=1m$),   $\eta$ is the Kolmogorov length scale and it is the smallest hydrodynamic scale in turbulent flows. 


The turbulence model used in this test is the $k -\omega$ SST model, which in several works (see for instance \cite{PIGAZZINI2018299}) proved capable of simulating turbulent flows associated with vortex induced vibrations.
 The Implicit Euler scheme is used for time discretization. 
 As for the spatial gradients, a Gauss linear scheme is employed. The convective and diffusive terms have been approximated with the first-order Upwind scheme \cite{Spalding1972} for more stability. The reason is that, in the transport-dominated turbulent regime here under study, the local Peclet number can reach peak values greater than 2.
The values of the relaxation factors $\alpha_u$, and $\alpha_p$ are fixed at 0.7 and 0.3 respectively. One non-orthogonal correction at each PIMPLE iteration is used to deal with the mesh's non-orthogonality. In addition, one pressure correction (inner correctors) and two momentum corrections (outer correctors) are used in the simulations. The linear solver selected combines a smoother Gauss-Seidel solver used for the pressure equation, and a symmetric Gauss-Seidel solver for the momentum equation. 
The structural motion is computed by means of the \texttt{sixDoFRigidBodyMotionSolver} OpenFOAM solver. Given the external fluid dynamic forces acting on a rigid body, such a solver is able to compute the linear and angular displacements in three dimensions. However, the airfoil here considered is only free to translate along the vertical direction (plunge displacement) and rotate along the axis perpendicular to the planar domain $\Omega$ (pitch displacement).
The resulting system of two second-order differential \cref{structmotion1,structmotion2} is solved using the Symplectic second-order explicit time-integrator for solid-body motion \cite{dullweber1997symplectic}. 
The  Arbitrary Lagrangian-Eulerian (ALE) method deals with the motion of the grid nodes resulting from the fluid-structure coupling. 
In particular, the plunging displacement $h$, and the pitching displacement $\theta$ of the airfoil are used to deform the mesh (in the transverse and rotational directions), as described in Sub\cref{pod-rbf}. Table \ref{tab:simparams} reports a comprehensive summary of all the modeling and numerical parameter values used for the simulation setup.
\subsection{Simulation results}
This section presents the results obtained with the reduced model developed on the airfoil test case described earlier.
As mentioned, the ROM is based on the POD-Galerkin approach for the momentum and continuity equations (velocity and pressure fields), on POD-LSTM or POD-NNs for the online eddy viscosity computation, and POD-RBF for the mesh displacement update. 
\begin{table}[H]
\caption{Summary of simulation settings of the flow passing pitch-plunge airfoil}
\label{tab:simparams}
\centering
\begin{tabular}{|p{3.25cm}|p{3.25cm}|p{3.25cm}|p{3.25cm}|}
\hline
\multicolumn{2}{|c|}{\textbf{Flow settings}} & \multicolumn{2}{|c|}{\textbf{Structure settings}} \\
\hline
Re &  $10^7$ & $f_{sh} = f_h = f_{\theta}$ & $\SI{20}{\hertz}$ \\
\hline
Time scheme & Implicit Euler & Time scheme & Sympletic\\
\hline
Gradient scheme   & cellLimited Gauss linear 1& $m$   &   $\SI{22.9}{\gram}$\\
\hline
Convective scheme&  Gauss upwind & $g_z$   &  $\SI{-9.81}{\metre\per \s^2}$\\
\hline
Laplacian scheme & Gauss linear limited 0.5 & $L_z$ & -2\\
\hline
$St = \dfrac{f_{sh}c\sin{\alpha}}{U_{\infty}}$    & 0.2 & $k_h$ & $\SI{3.6262e+5 }{\newton \per \metre}$ \\
\hline
$U_{in}$ &   $\SI{e+2}{\metre\per \s}$  &  $k_{\theta}$ & $\SI{3.25e+4}{\newton \per \metre}$\\
\hline
Co  & 0.5  & $c_h$   &  $\SI{2}{\newton \per \metre}$\\
\hline
${\Delta t}_{DNS} $ & $ \frac{C_0\times Re^{-\frac{3}{4}}}{U_{in}}$  & $c_{\theta}$   &  $\SI{5e-1}{\newton \per \metre}$\\
\hline
Turbulence model & $k -\omega$ SST  & $I_{zz}$   & 2.057121362\\
\hline
\end{tabular}
\end{table}
In Table \ref{tab:simparams}, $St$ is the Strouhal number, $f_{sh}$ the frequency of the vortex shedding. $L_z$ is the angular momentum in Z-direction,  and $g_z$ the gravity in the Z-direction.
The following quantities: $m$, $f_h$, $f_{\theta}$, $k_h$, $c_{\theta}$,  $k_{\theta}$, $c_h$, and  $I_{zz}$ are defined in \cref{structmotion}.  


Note that the FVM C++ library ${OpenFOAM}^{®}$ version 2106 \cite{jasak2007openfoam} has been used for data collection at the full-order model level. Such a numerical solver, widely used in industrial applications \cite{Jasak2009, shademan2013evaluation} exploits the fact that FVM locally respects the balance of momentum and mass.  
At the reduced level, the reduction and resolution of the reduced system are carried out using the C++-based library ITHACA-FV (In real Time Highly Advanced Computational Applications for Finite Volumes) \cite{stabile2018finite, StabileHijaziMolaLorenziRozza2017}. ITHACA-FV is designed to carry out Galerkin projection of PDE problems that, at the full-order level, are solved making use of FV discretization based on OpenFOAM. The interpolation using RBF in this work has been carried out using the C++ library SPLINTER \cite{SPLINTER}. The radial basis used for interpolation is the \textit{thin plate spline} with the radial basis's radius set to 1. Finally, the RNNs/ NNs are built using the PyTorch library \cite{NEURIPS2019_9015}. The next Subsection assesses the qualitative prediction of the reduced model.


\subsubsection{Prediction quality}
\begin{figure}[H]
\centering
\includegraphics[width=\textwidth]{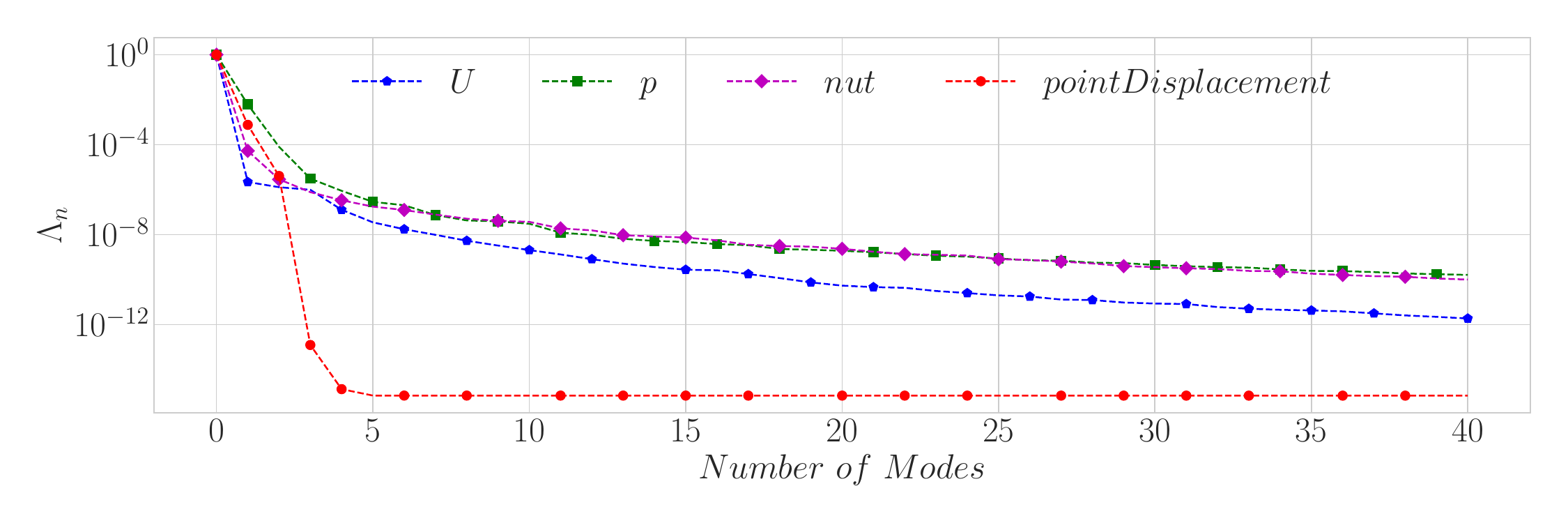}
    \caption{The decay of the POD modes eigenvalues for velocity, pressure, pointDisplacement, and Eddy viscosity fields. Color code: blue -- velocity, green --pressure, red -- pointDisplacement, magenta -- eddy viscosity}
\label{fig:eigdecay}
\end{figure}

The point of this Subsection is that of establishing the accuracy of the modal decomposition upon which the reduced model is based.
Table \ref{tab:eigenvectors} presents the eigenvalues associated with the first five dominant modes of all the fields of interest. 
The values reported also suggest that for the grid nodes motion (pointDisplacement field), one single mode could be enough to predict the mesh motion with acceptable accuracy. Hence, in this study, 1 mode will be used to predict the airfoil motion. A more comprehensive view of the modes eigenvalues magnitude is presented in \cref{fig:eigdecay}.


\begin{table}[H]
\centering
\begin{tabular}{|p{1.25cm}|p{2.75cm}|p{2.75cm}|p{2.95cm}|p{3cm}|}
\hline\noalign{\smallskip}
\# Modes & Eigenvalues U &  Eigenvalues p &   Eigenvalues $\nu_t$ &         Eigenvalues pointDisplacement\\
\noalign{\smallskip}\hline\noalign{\smallskip}
1 & 1           & 1                        & 1                             &  1 \\
2 & 0.001480623 & 0.07835354               & 0.007290981                   & 0.027573111 \\
3 & 0.001133095 & 0.008968419              & 0.001695786                   & 0.00201184\\
4 & 0.000992664 & 0.001758634              &  0.000879206                  & 0.000000351\\
5 & 0.000356768 & 0.000941899              & 0.000580963                   & 0.00000000018\\
\noalign{\smallskip}\hline
\end{tabular}
\caption{Normalized eigenvalues of the POD modes of the fields of interest}
\label{tab:eigenvectors}
\end{table}


\subsubsection{Machine learning of the temporal eddy viscosity coefficients}
For all the training runs in this work, a variant of the stochastic gradient descent algorithm called adaptive moment estimation ADAM \cite{kingma2017adam} is used as an optimizer. ADAM has an adaptive learning rate method which is commonly used to train deep networks. The optimization is based on a scaled version of the modal coefficients, given by

\begin{equation}
    \hat{\boldsymbol{a}}_j(t) = \frac{\boldsymbol{a}_j(t) - \left<\boldsymbol{a}_j(t)\right>}{\sigma[\boldsymbol{a}_j(t)]},
\end{equation} 
where $\left<\boldsymbol{a}_j(t)\right>$ is the mean value of the modal coefficient time series considered and $\sigma[\boldsymbol{a}_j(t)]$ is the corresponding variance.
The dataset scaling is necessary to avoid that the gradients that enter the computations of the cost function are too small. In such a case, it would be impossible to generate significant updates of the parameters of the network \cite{borrelli2022predicting}. The model parameters (weights and bias) are trained with the PyTorch library  \cite{paszke2019pytorch} and later imported in the C++ solver to generate the transient predicted solution for the eddy viscosity during the online computations. 
The full details on training, validation, and testing are reported in \ref{appendix}.
The accuracy of the resulting feed-forward NN and LSTM-RNNs model results are illustrated in Fig. \ref{fig:lstm-ffnns}. The four diagrams represent the time series of the modal coefficients corresponding to the four energetically dominant modes of the full-order model eddy viscosity, as well as their data driven approximations. In the diagrams, the two dashed vertical lines divide the time axis into the training window, on the left, the validation window, located between the red and blue dashed lines, and the testing window, on the right.
The picture confirms that both models are able to capture the overall trend of the reference FOM solution, not only in the training time window, but also in the and validation and testing ones. By a quantitative perspective, both data driven models appear accurate in the reproduction of frequency, amplitude and phase of the first three modal coefficients of the eddy viscosity field. The plot corresponding to the fourth modal coefficient shows instead a drop of the accuracy of the feed-forward NN prediction, while, on the other hand, the LSTM-RNN prediction remains as accurate as for the previous modes.  
Thus, this preliminary analysis suggests that both feed-forward NN and LSTM-RNN  data driven algorithms used are in principle capable of approximating with good accuracy the eddy viscosity modal coefficients. This is of course crucial for a correct closure of the turbulent problem at the reduced level. The next sections will then assess if the quality of the eddy viscosity approximation --- which appears high except for higher order modal coefficients obtained with feed-forward NN --- will translate into accurate ROM results.

\subsubsection{ROM online resolution time}
The data were generated by transient simulations running for one second and saving snapshots of the flow field every 0.0005 seconds for a total of 2001 snapshots. 
All simulations were run on an HP Pavilion laptop with  AMD Ryzen 7 5700u with Radeon graphics $\times 16$, 16GB RAM, AMD Renoir graphics card, and Ubuntu 20.04 operating system.
Table \ref{tab:OffOnline} reports a comparison analysis of the the full-order and reduced-order models execution times as the number of modes for the prediction of velocity, pressure, pointDisplacement, and eddy viscosity is varied. This allows for evaluating the effect of the number of modes variation on the computational cost of the online phase.
\begin{table}
\centering
\begin{tabular}{|p{4cm}||p{5cm}||p{4cm}| }
	\hline
	\textbf{Stages} & \textit{\# of modes} & Time [s]\\
 \hline
	\textbf{Offline} &  - & 3.441516e+4\\
 \hline
    \multirow{3}{*}{\textbf{POD-NNs}} & $N_u=N_p=5$, $N_{nut}=3 $, $N_{pD} = 1$ & 1.770725259e+4\\
       \cline{2-3}
       &    $N_u=N_p=10$,$N_{nut}=5$,$N_{pD}=1$ & 1.957406359e+4\\
       \cline{2-3}
       & $N_u= 15 , N_p=5$, $N_{nut}=2$, $N_{pD}=3$ &  2.159144848e+4 \\
       \cline{2-3}
       &$N_u =10,N_p =5,N_{nut}=3$, \textit{and} $N_{pD}=1$ &   1.942751287e+4\\
   \hline
    {\textbf{POD-LSTM}} & $N_u =N_p=N_{nut}=5 $ \textit{and} $N_{pD} = 3$ & 1.766056714e+4\\
    \cline{2-3}
    & 
    $N_u =15,  N_p = 5$,  $N_{nut} = 3 $ \textit{and} $N_{pD} = 3$ & 2.136608204e+4\\
       \hline
\end{tabular}
\caption{Offline and Online times comparison varying the number of modes}
\label{tab:OffOnline}
\end{table}

The offline stage comprises four steps: the computation of the snapshot (computed by a numerical approximation of
the original high-dimensional system),  computation of the POD basis, projection of the dynamics on the low-rank subspace, and the Machine Learning training of the neural networks (including for the radial basis networks). But only the computational coast of the first step is  reported in Table \ref{tab:OffOnline} as it is the most expensive one. The online coast is the computational time needed to compute the solutions of the surrogate model. In Table \ref{tab:OffOnline},  one can observe a small speed-up as this work does not employ hyper-reduction technique.

However, the results in  Table \ref{tab:OffOnline} suggest that the speed-up obtained by the online solution of the reduced system is not proportional to the reduction of the unknowns  obtained at the reduced-order level. This is due to the fact that in the presence of a deforming domain such as the one characterizing our FSI simulations, the entries of the matrices of the ROM system must be computed at each time step through integrals on the updated full-order grid. Clearly, this is at the moment representing a major bottleneck towards a ROM which grants significant computational cost reduction with respect to its FOM counterpart, and work is being carried out towards lowering the computational cost associated with the reduced model assembling. Nonetheless, the main goal of the present work is that of assessing the accuracy of the ROM approach taken. In particular, it is important establishing whether the interaction between the physics based reduction of the fluid dynamic balance equations, and the data driven reduction of the turbulence and grid displacement equations, results in an accurate solver. 

\subsubsection{ROM online resolution quality}

The present Subsection aims at analyzing how close the predicted ROM solutions are with respect to the FOM ones. To this end,  \cref{fig:fieldofinterest1,fig:fieldofinterest2} shows a qualitative comparison between the solution fields contour plots corresponding to time $t=0.1\ $s obtained with both the FOM and ROM solvers. The plots in the figure confirm that, to the eyeball test, the ROM solutions obtained using both POD-NN and POD-LSTM appear by all means similar to the high-fidelity ones.

\begingroup
\setlength{\tabcolsep}{0.05pt} 
\renewcommand{\arraystretch}{0.05} 
\begin{figure}
\centering
\begin{tabular}{ccc}
\includegraphics[width=0.285\linewidth]{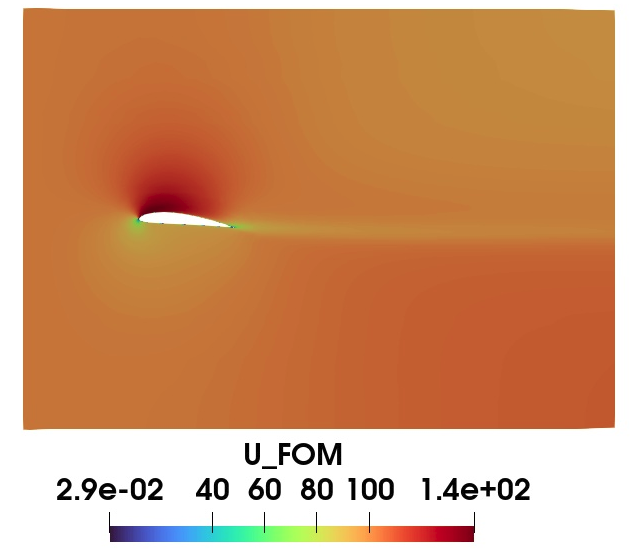} &  \includegraphics[width=0.285\linewidth]{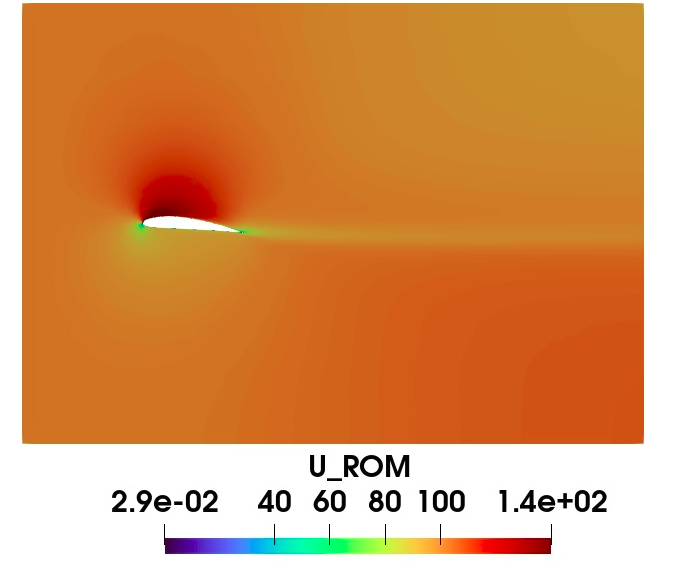} &
\includegraphics[width=0.265\linewidth]{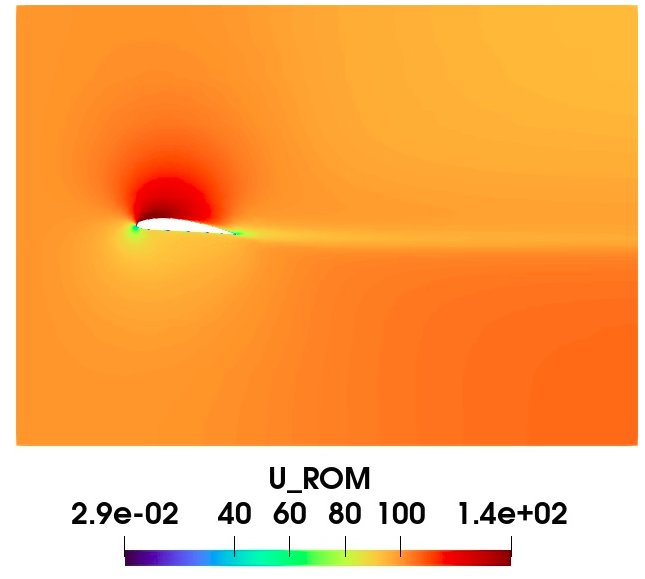}\\
\includegraphics[width=0.28\linewidth]{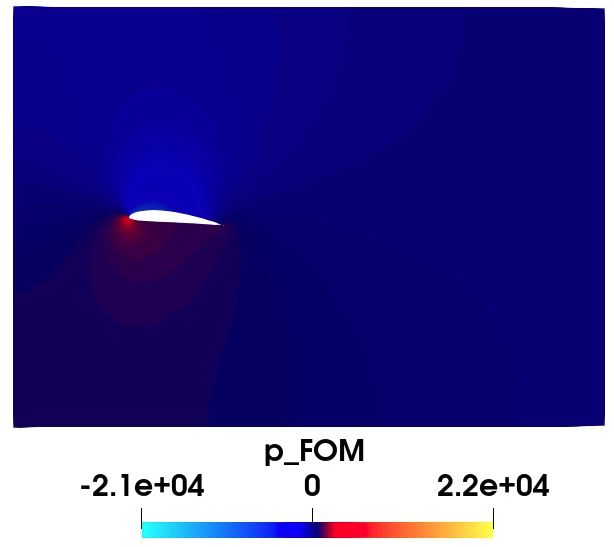} &  \includegraphics[width=0.28\linewidth]{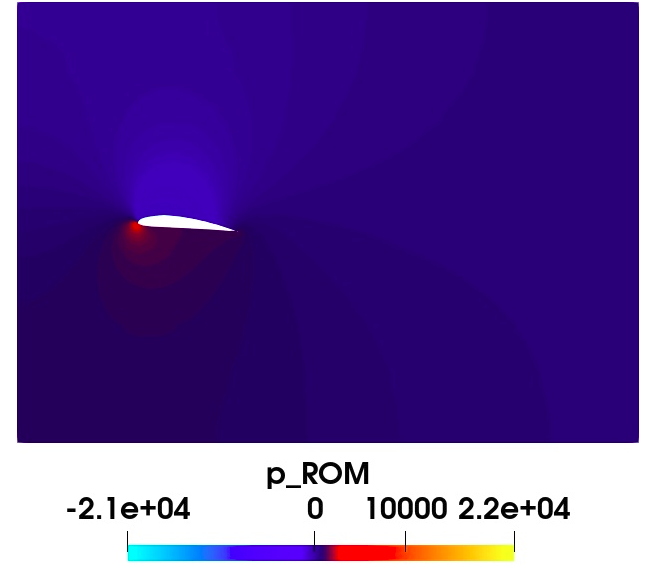} &
\includegraphics[width=0.28\linewidth]{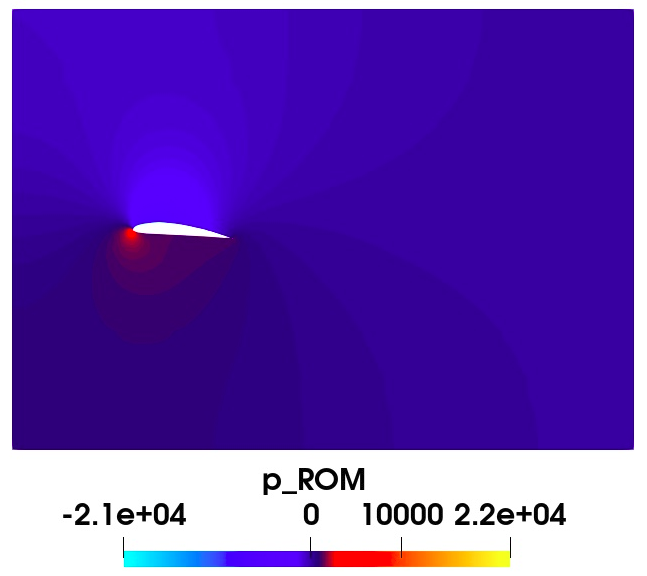}\\
(a) & (b)  &  (c)
\end{tabular}
\caption{Comparison of the velocity and pressure fields.  First row velocities comparison and  second row pressure comparison. Column (a) FOM fields, column (b) reduced solution with POD-NNs and column (c) reduced solution with POD-LSTM.
The snapshots are captured in the second period i.e. $t=T=0.1\ $s}	
\label{fig:fieldofinterest1}
\end{figure}
\endgroup

\begingroup
\setlength{\tabcolsep}{0.05pt} 
\begin{figure}
\centering
\begin{tabular}{ccc}
\includegraphics[width=0.285\linewidth]{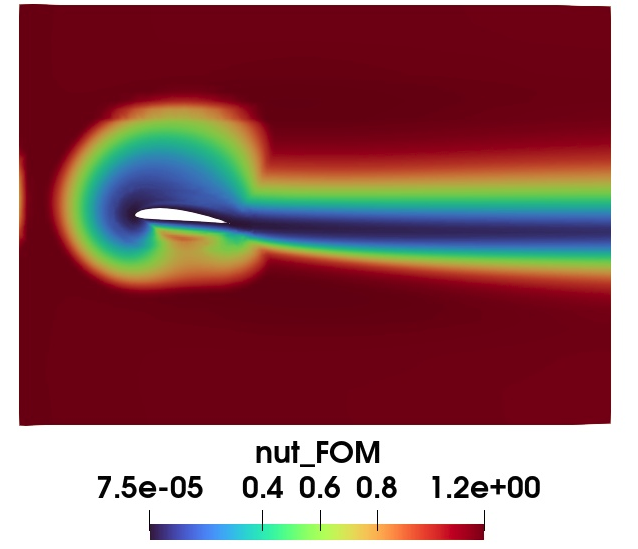} &  \includegraphics[width=0.27\linewidth]{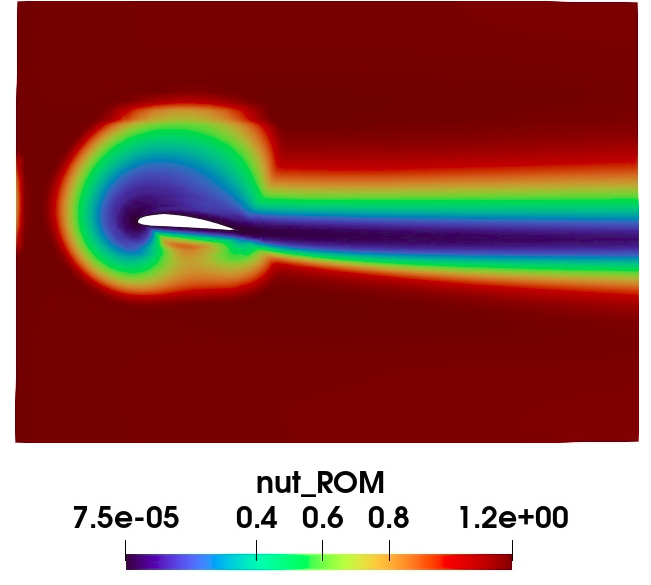} &
\includegraphics[width=0.27\linewidth]{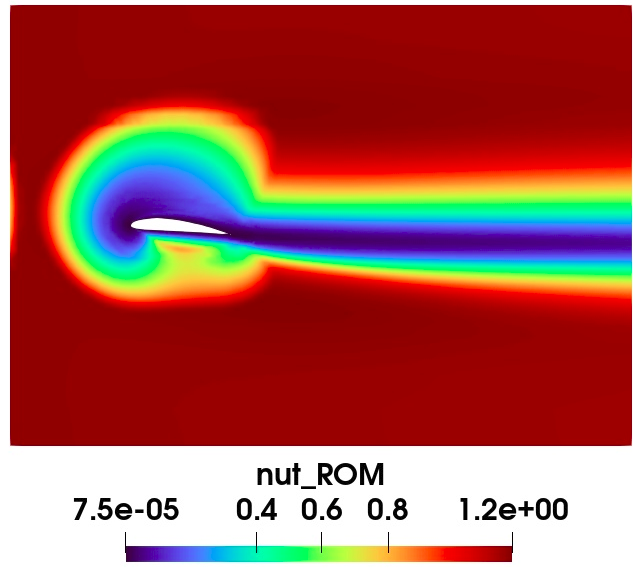}\\
\includegraphics[width=0.275\linewidth]{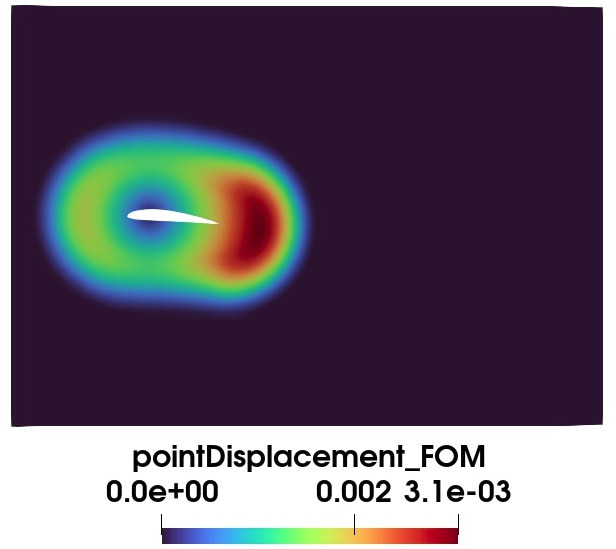} &  \includegraphics[width=0.275\linewidth]{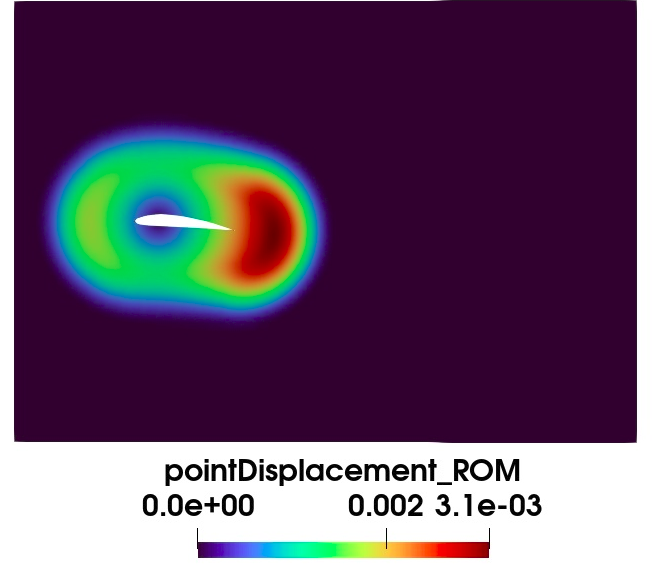} &
\includegraphics[width=0.275\linewidth]{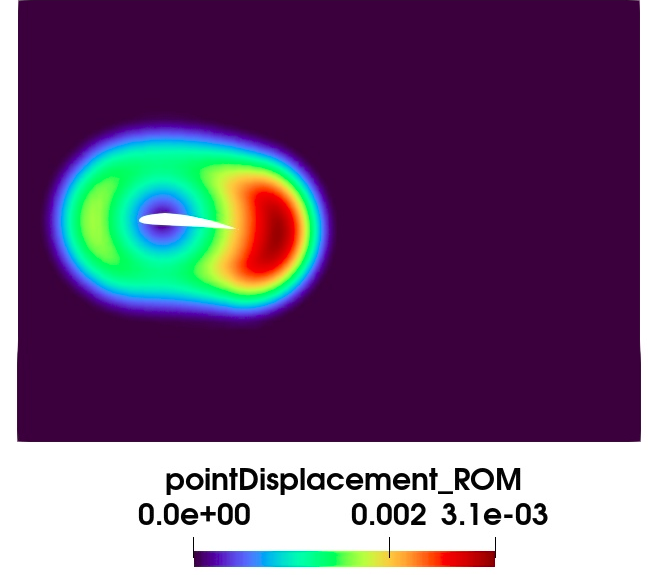}\\
(a) & (b)  &  (c)
\end{tabular}
\caption{Comparison of the eddy viscosity and grid node displacement fields.  First row eddy viscosity comparison and  second row grid node displacement comparison. Column (a) FOM fields, column (b) reduced solution with POD-NNs and column (c) reduced solution with POD-LSTM.
The snapshots are captured in the second period i.e. $t=T=0.1\ $s}	
 \label{fig:fieldofinterest2}
\end{figure}
\endgroup

\begin{figure}
\centering
\includegraphics[width=1\textwidth]{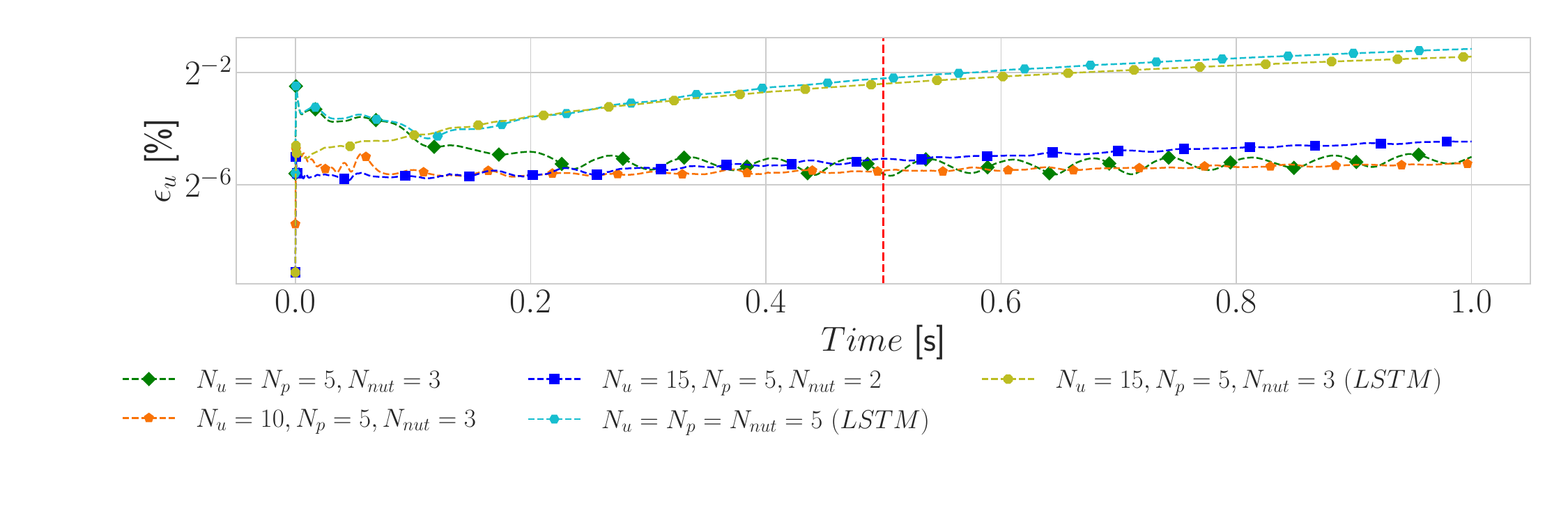}
\caption{Sensitivity study of the error (log-scale) in the  $L^2$-norm  versus the time evolution of the velocity field. The red line shows the results obtained inside and outside the time window}
\label{fig:errorsvelocity}
\end{figure}
\begin{figure}
\centering
\includegraphics[width=1\textwidth]{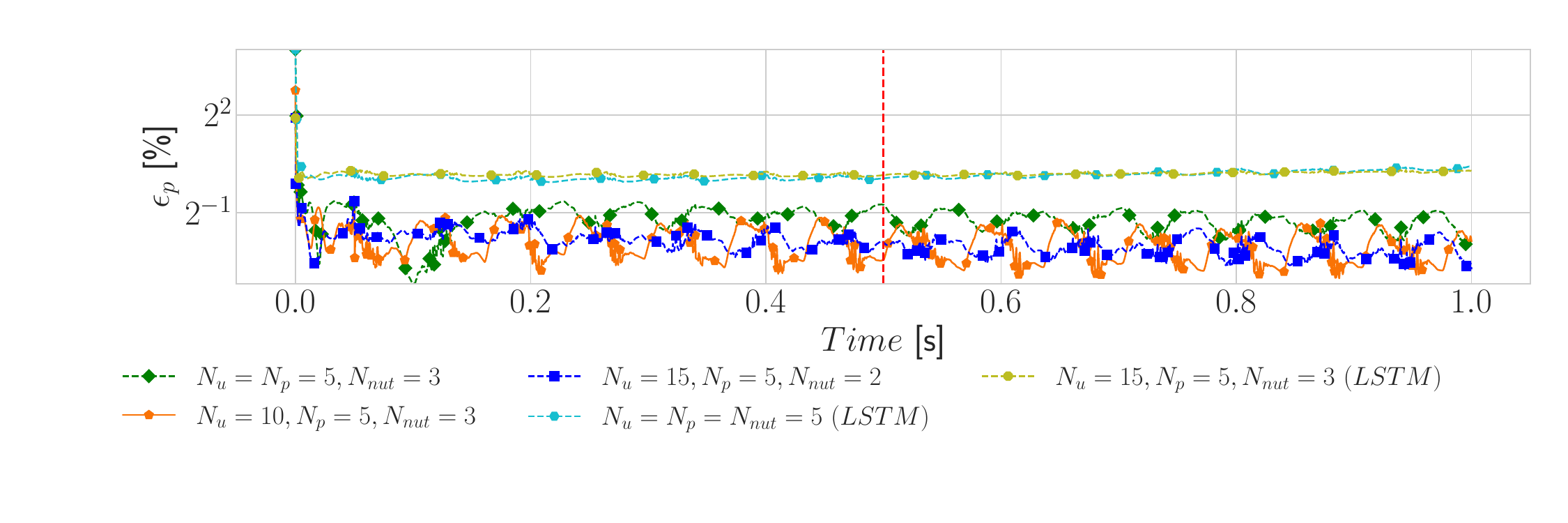}
\caption{Sensitivity study of the error (log-scale) in the  $L^2$-norm versus the time evolution of the pressure field. The red line shows the results obtained inside and outside the time window}
\label{fig:errorspressure}
\end{figure}
\begin{figure}
\centering
\includegraphics[width=1\textwidth]{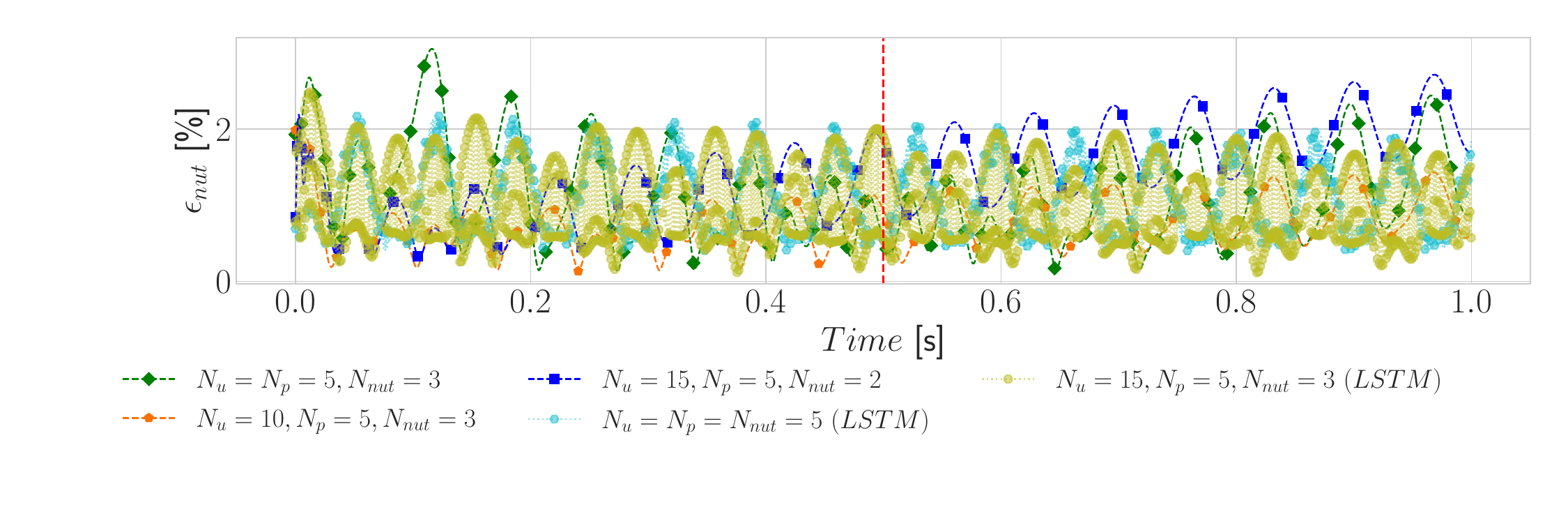}
\caption{Sensitivity study of the error in the  $L^2$-norm in log-scale versus the time evolution of the eddy viscosity field. The red line shows the results obtained inside and outside the time window}
\label{fig:errorsNut}
\end{figure}

More quantitative considerations can be driven from \cref{fig:errorsvelocity}, \cref{fig:errorspressure}, and \cref{fig:errorsNut} which depict the time evolution of the ROM $L^2$ error of the velocity, pressure, and eddy viscosity obtained with different combinations of modal truncation orders for the velocity field. Note that the  $L^2$ relative error for a given quantity $q$ is computed as follows:
\begin{align}
\epsilon_q = \displaystyle\frac{||q_{FOM} - q_{ROM} ||_{L^2(\Omega(t)  )}}{||q_{FOM}||_{L^2(\Omega(t)   )}}\times 100\%.
\end{align}
The results in \cref{fig:errorsvelocity} confirm the qualitative impression of accuracy given by \cref{fig:fieldofinterest1,fig:fieldofinterest2}, as the velocity field errors plotted remain well below the 1\% threshold for the entire simulation. The plot also clearly indicates that the velocity field accuracy obtained making use of POD-NN is higher than that obtained with the POD-LSTM approach. This is further confirmed by \cref{fig:errorspressure}, in which the error associated with POD-LSTM is approximately as high as 1\%, while the POD-NN approach results in appreciably lower error levels. The satisfactory results shown in these plots depend on the NN / LSTM algorithm effectiveness in the calculation of the  eddy viscosity field POD coefficients time evolution. In this regard, it is worth pointing out that the \cref{fig:errorsNut} error magnitude suggests that a 1-2\% error level for the eddy viscosity field approximation, leads to consistently lower error values on the velocity and pressure fields. 
These observations seem to further validate the data driven approach taken for the eddy viscosity field, as it appears to lead to small errors of pressure and velocity fields, which are the main fields of interest for our simulations. Fig. \ref{fig:errorsNut} also shows that the error associated with the POD-LSTM approach presents a pattern characterized high frequency oscillations, which might be the culprit for the higher error levels observed in the velocity and pressure fields. This is consistent with what other researchers have observed in \cite{borrelli2022predicting,srinivasan2018deep,Eivazi2021} in the prediction of the velocity field in a channel flow. In their study, they reported that the phenomenon was due to an insufficient number of snapshots of the training dataset or when the number of cells in the hidden layer was not enough. In addition to that, the lifetime of a transient turbulent state is highly sensitive to the initial conditions and if  only one component of the initial conditions differs by $10^{-12}$, the resulting trajectory will diverge from the truth one. 
It is also possible that the memory in the sequence can affect the LSTM accuracy as reported in \cite{mohan2018deep} because signals originating from chaotic dynamic systems are known to have quite short correlated events and memory does not typically persist over long periods. In this case, the Hurst exponent is a prominent solution \cite{hurst1951long} for further investigations. However, the mentioned suggestions were not the emphasis of this paper. Instead, the emphasis was concentrated on the design of the overall solver to generate a reduced model for segregated FSI solvers for turbulent regime in the FV context.

The plots in \cref{fig:errorsvelocity}, \cref{fig:errorspressure}, and \cref{fig:errorsNut} also visualize the effect of the number of velocity modes on the accuracy of the ROM solutions. In general, increasing the number of velocity modes considered at the online level increases the accuracy, especially in the initial transient part of the time integration. However, an increase to values higher than 10 modes does not appear to result in significant gains. It is finally important to point out that the snapshots for the POD have been collected only in a training window corresponding to the first half of the time history plotted --- the one on the left of the red dashed vertical line in each plot. So, the plots also indicate that, in the case of a periodic problem as the one analyzed, the ROM errors are not significantly growing if time extrapolation is carried out. 

A further step in the ROM results analysis is represented by the evaluation of the fluid dynamic forces and airfoil displacement accuracy. 
In fact, the $L_2$ errors discussed in the previous plots provide information on the average discrepancy of the most relevant fluid dynamic quantities in the fluid domain. However, the previous plots provide little information on the local distribution of such errors, which might have relevant impact on our FSI simulations. In fact, a low overall error in the pressure and velocity fields might still be in principle associated with high local error in small regions, for instance surrounding the airfoil. In such a case, both the fluid dynamic forces and the airfoil displacement might be computed with low accuracy.

Fig. \ref{fig:lift} depicts the time history of the lift force exerted by the fluid on the airfoil. In the plot, the FOM solution is compared to the ones obtained with ROMs making use of different number of pressure and velocity modes. The plot suggests that both the ROM methodologies proposed are able to obtain qualitatively good approximation of the lift force throughout the time integration window considered, even including the initial transient. Also in this case, the plot indicates with a red dashed vertical line the separation between the training window, on the left, and the time window, on the right, in which the solution is extrapolating over the time variable. An inspection of the lift curves suggests that, when POD-NN is considered, no significant error increase is associated to time extrapolation. As for POD-LSTM, a slight degradation of the prediction quality is observed in the extrapolation region. Further confirmation of this is given by the corresponding error plots presented in \cref{fig:errorlift}, in which it is possible to observe that for all the combination of pressure and velocity modes considered, the error remains below the 1\% threshold, except for the initial transient, in which the $N_u=N_p=5$  solution for both POD-NN and POD-LSTM presents a slightly higher error.

\begingroup
\begin{figure}
\centering
\includegraphics[width=\textwidth]{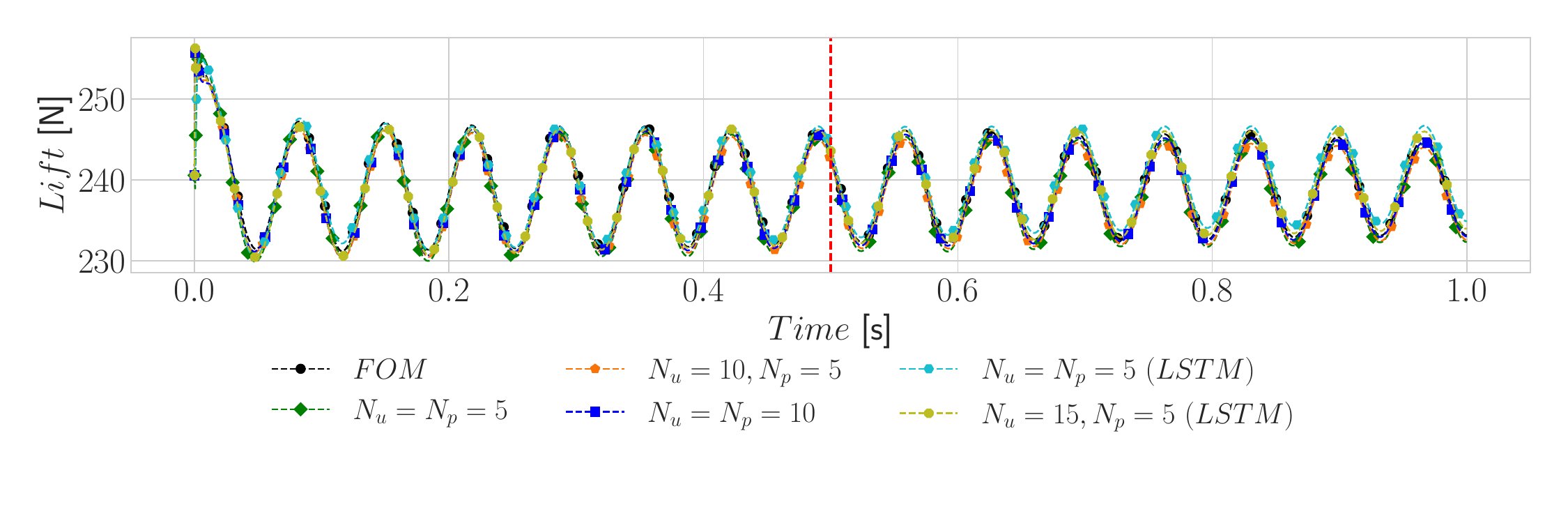} 
\caption{Times series comparison between the reference signal of the lift force acting on the foil in Newton unit with predicted signals. The vertical line in red divides the signal in two: left prediction in the time window and right prediction outside the time window (extrapolation)}
\label{fig:lift}
\end{figure}

\begin{figure}
\centering
\includegraphics[width=1\textwidth]{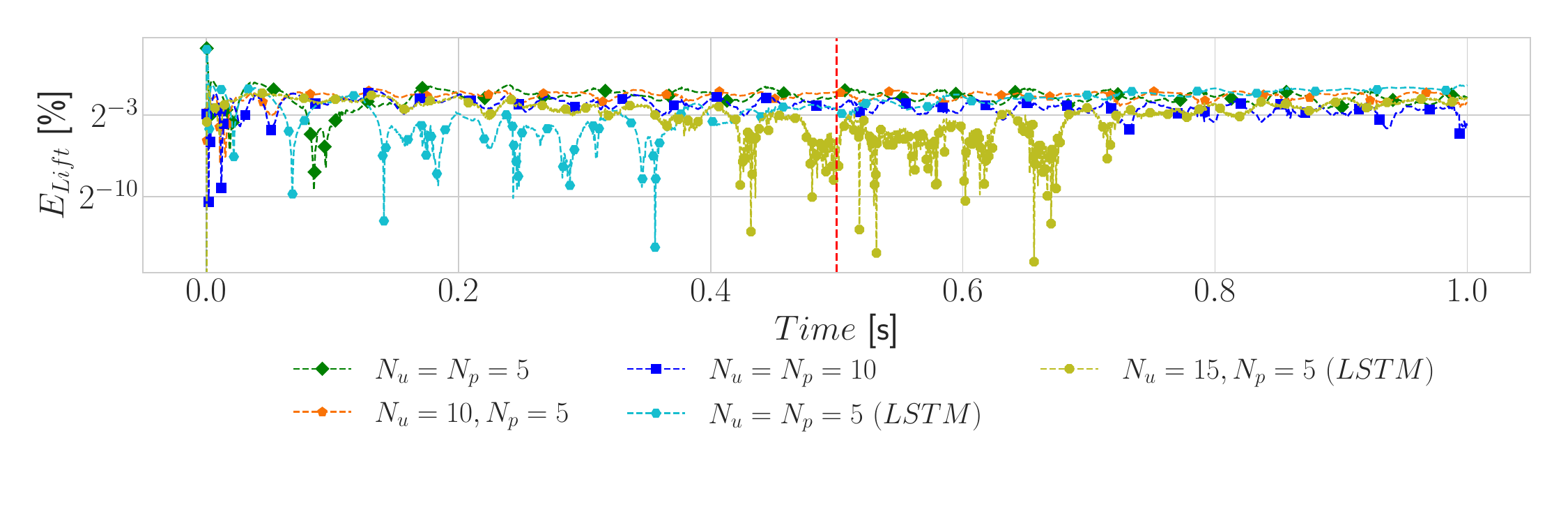}
\caption{Times series of the error analysis of the lift force (original and predicted signals) from \cref{fig:lift}. The vertical line in red divides the error plots in two. Left: interpolation error and right: extrapolation error}
\label{fig:errorlift}
\end{figure}

Similar plots relative to the airfoil drag are presented in \cref{fig:drag} and \cref{fig:errordrag}. Also in this case, the figure presents a comparison between the FOM drag curve and the corresponding curves obtained with ROM models making use of different modal truncation orders. The plot suggests that the qualitative behavior of the airfoil resistance is well captured across all time steps of the flow simulation, including the extrapolation time window. The value of the drag error obtained with POD-NN is again lower than 1\% for the most part of the overall time history, although non-negligible portions of the error curve pass such a threshold. Such a higher relative error is just a product of the lower absolute value of the drag force with respect to the lift values. Thus, it can be said that the accuracy shown by the plots is quite remarkable, and is not degrading in the time extrapolation window. On the other hand, the error obtained with POD-LSTM appears to settle for higher values, although it remains below 5\% for the time window analyzed, including the extrapolation region. Also, in this case, the increased error scale with respect to the lift should be associated with the smaller magnitude of the drag force absolute value and amplitude. 

\begin{figure}
\centering
\includegraphics[width=1\textwidth]{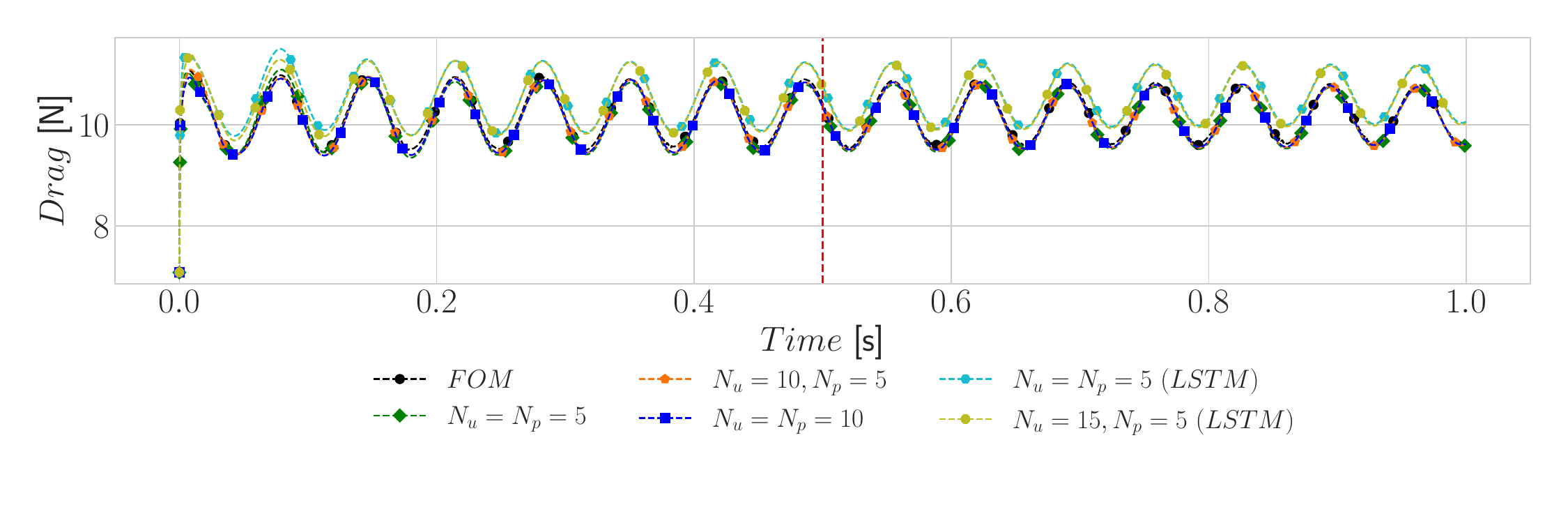} 
\caption{Times series comparison between the reference signal of the drag force acting on the foil in Newton unit with predicted signals. The vertical line in red divides the signal in two: left prediction in the time window and right prediction outside the time window (extrapolation)}
\label{fig:drag}
\end{figure}
\begin{figure}
\centering
\includegraphics[width=1\textwidth]{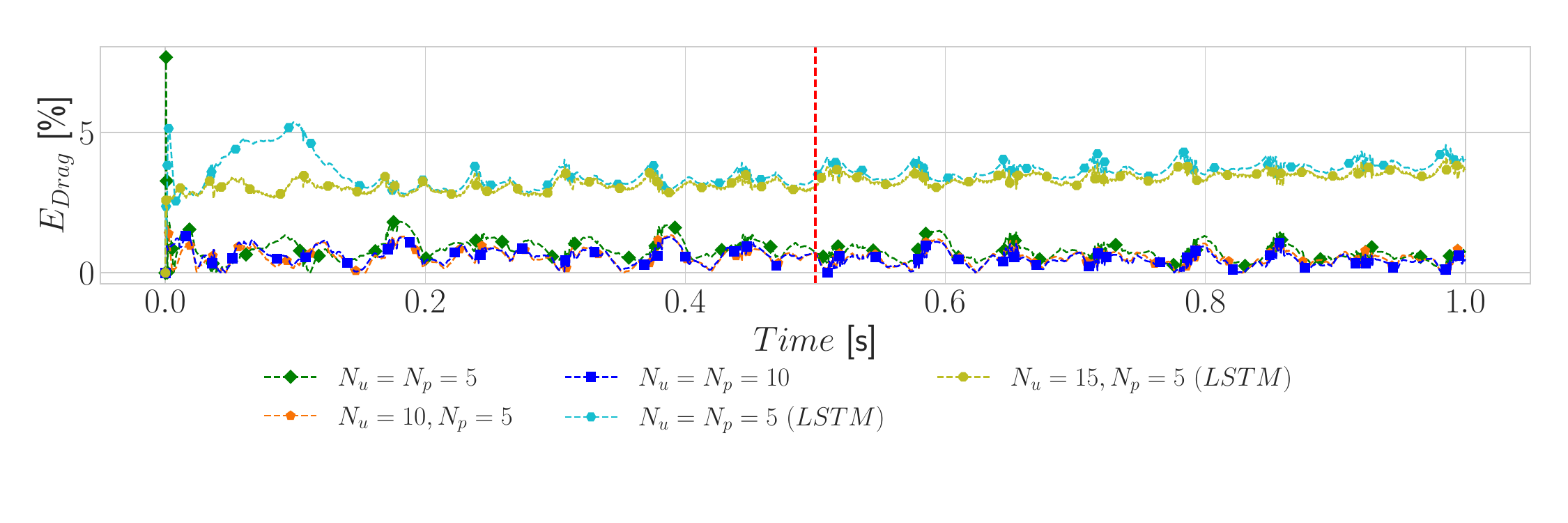}
\caption{Times series of the error analysis of the drag force (original and predicted signals) from \cref{fig:drag}. The vertical line in red divides the error plots in two. Left: interpolation error and right: extrapolation error}
\label{fig:errordrag}
\end{figure}

As a final confirmation of the proposed ROM results quality, it is important to also consider the time history of the airfoil displacements computed during the simulations. Fig. \ref{fig:CentreOfRotation} depicts the airfoil center of gravity position, as computed at each time step by the FOM and by the ROM models proposed. The plot clearly indicates that all the reduced-order model solutions closely track the full-order one. For the most part of the time window --- which as usual is divided into a training part and an extrapolation part by the red dashed line in the plot --- the plunge coordinates curves obtained with all the POD-NN models considered appear in fact overlapped to the FOM one. The POD-LSTM results obtained making use of $N_u=15$ velocity modes and $N_p=5$ pressure modes have accuracy comparable with the POD-NN ones. Instead, the POD-LSTM curve associated with $N_u=N_p=5$ visually appears less accurate, especially in the extrapolation region. The corresponding error plot is presented in \cref{fig:errorplungd}. The values reported in the diagram substantially confirm that the POD-NN error obtained with all the modal truncation combination considered, is in average as low as 0.1\%, and always below the 1\% threshold. Again, it has to be remarked that this satisfactory result does not appear to be negatively affected by time extrapolation, as the values in the second half of the plot remain generally low.

\begin{figure}
\centering
\includegraphics[width=1\textwidth]{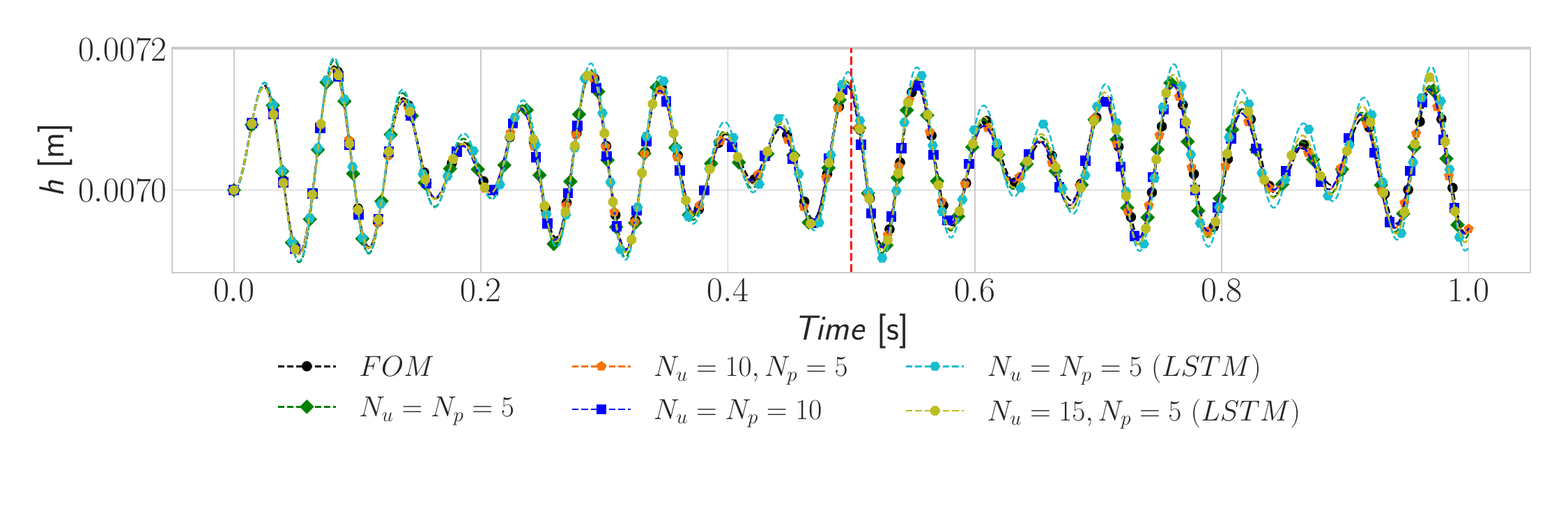}
\caption{Time series comparison between the reference signal of the plunge with different predicted signals. The vertical line divides the signal in two: left prediction in the time window and right prediction outside the time window (extrapolation) }
\label{fig:CentreOfRotation}
\end{figure}

Fig. \ref{fig:AngleOfAttack} shows the time history of the airfoil pitch angle, as simulated with the FOM solver, and with ROM solvers making use of different eddy viscosity approximation strategies and different modal truncation orders. Here, all the ROM solvers tested --- even the ones with lower truncation orders --- lead to airfoil pitch curve approximation that are barely distinguishable from the FOM one. This is suggesting that not only the aerodynamic force, but also its point of application are reproduced with accuracy at the reduced order level. As a consequence, the pitch angle percentage error plot, presented in \cref{fig:errorpitch} displays errors that are consistently below the 0.1\% value throughout the entire simulation, for all the ROM models considered.

\begin{figure}
\centering
\includegraphics[width=\textwidth]{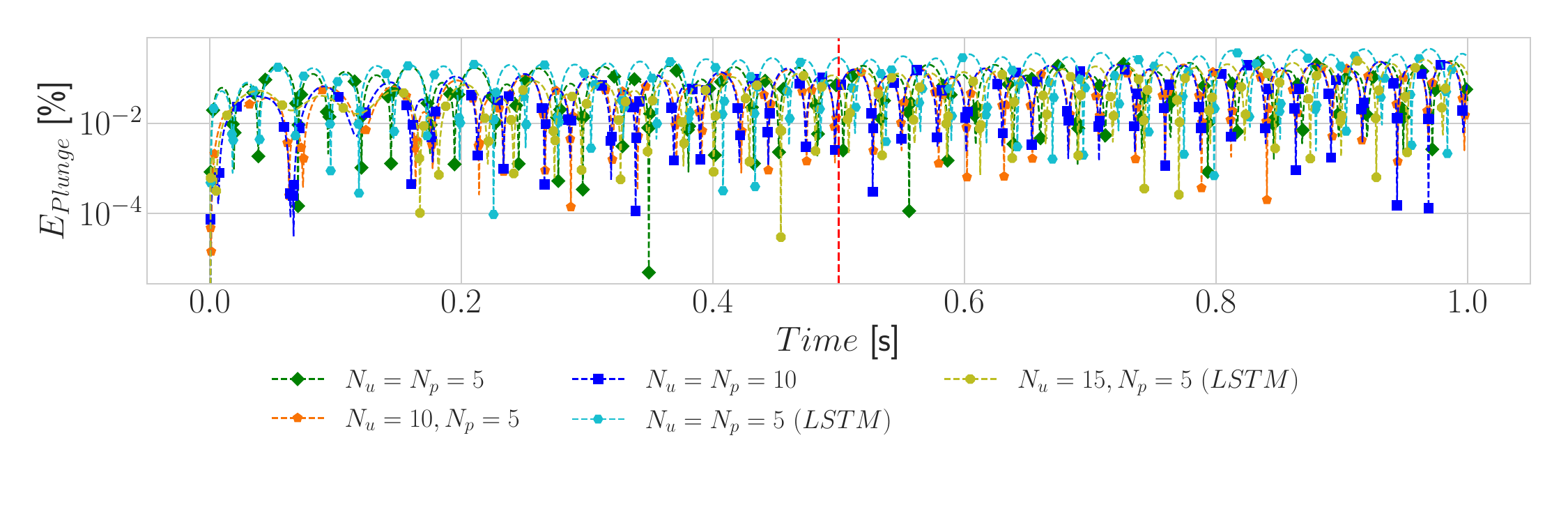}
\caption{Plunge's error analysis versus time}
\label{fig:errorplungd}
\end{figure}

\begin{figure}
\centering
\includegraphics[width=\textwidth]{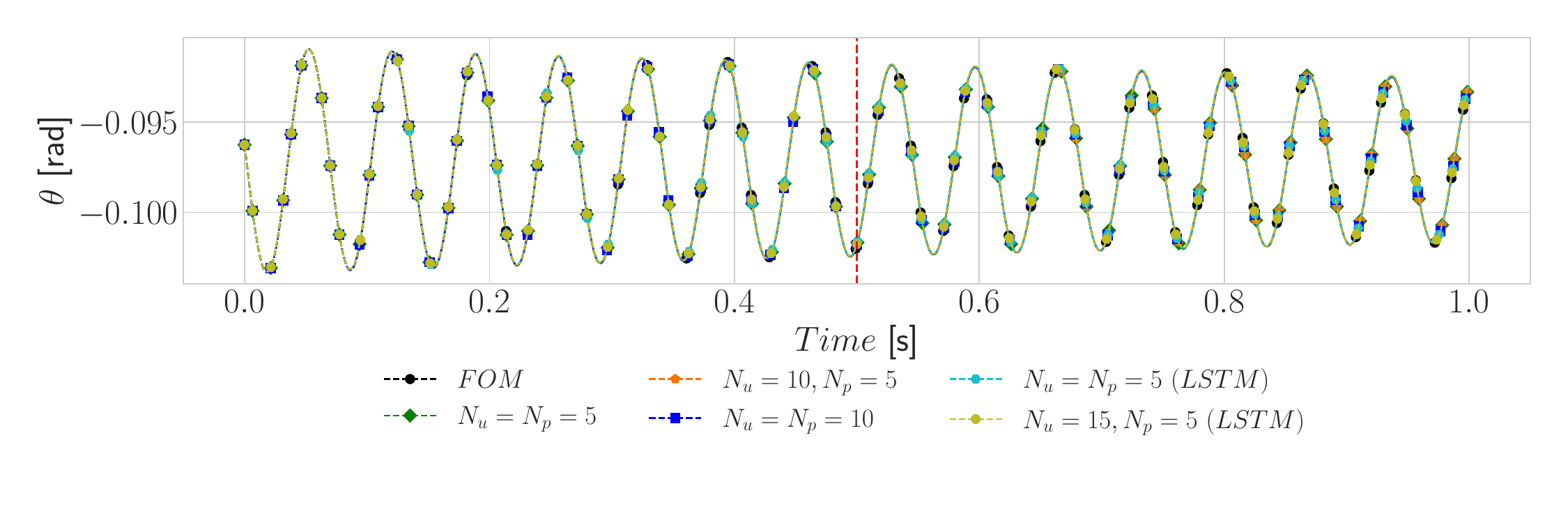}
\caption{Time series comparison between the reference signal of the pitch (angle of attack) with different predicted signals. The vertical line in red divides the signal in two: left prediction in the time window and right prediction outside the time window (extrapolation)}
\label{fig:AngleOfAttack}
\end{figure}

\begin{figure}
\centering
\includegraphics[width=\textwidth]{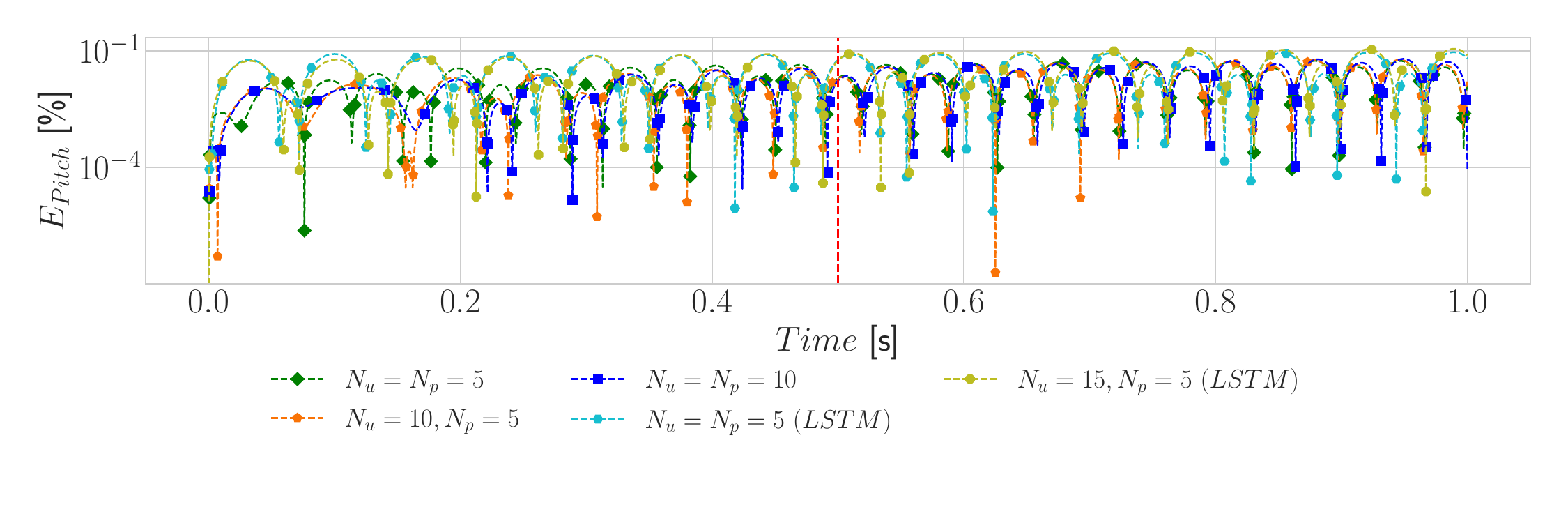}
\caption{Pitch's error analysis versus time}
\label{fig:errorpitch}
\end{figure}

%% file: sections/conclusion.tex
\section{Conclusions and perspectives}\label{conclusion}
This paper has proposed a hybrid projection-based ROM for segregated fluid-structure
interaction (FSI) solvers in an ALE approach at a very high Reynolds number.
The Finite Volume Method (FVM) has been used as discretization technique at the full-order level as the method ensures conversation properties and it is a preferred choice in industry for solving complex engineering problems as well as  its ability to handle real-world geometries. This method is designed to work effectively with segregated solvers within the ALE framework at very high Reynolds numbers.
 As traditional methods can be computationally expensive, a reduced method has been introduced that combines a POD-Galerkin projection, POD-RBF, POD-NNs, and POD-LSTM as these methodologies have demonstrated immense potential in other  research's fields. 
The resulting ROM framework proposed here is modular, hybrid, and data-driven; thus, it aligns with the development of a digital twin including fluid-structure
interaction effects.
The ROM is tested against a benchmark FSI problem at high Reynolds numbers to validate its accuracy and efficiency.
By comparing the proposed ROM to the full-order model, the results indicated that the proposed hybrid ROM achieved acceptable accuracy, stability, and convergence. 
The proposed hybrid projection-based ROM offers a promising solution for efficiently solving segregated FSI problems in an ALE framework at high Reynolds numbers without compromising accuracy.
This approach could be particularly valuable for industries dealing with complex FSI problems, such as aerospace and automotive engineering, where computational resources are a critical concern.
A natural direction for future work will include the construction of a robust LSTM encoder-decoder model with an attention layer to keep track of coefficient relationships within the inputs and output coefficients.  Another avenue for further improvement is to equip the standard Galerkin projection with a hyper-reduction algorithm to boil down the computation time.


%% file: sections/ackno.tex
\section*{Acknowledgements}
This work was partially funded by European Union Funding for Research and Innovation --- Horizon 2020 Program --- in
the framework of European Research Council Executive Agency: H2020 ERC CoG 2015 AROMA-CFD project 681447 ``Advanced Reduced Order Methods with
Applications in Computational Fluid Dynamics'' P.I. Professor Gianluigi Rozza, by PRIN ``Numerical Analysis for Full and Reduced Order Methods for Partial Differential Equations'' (NA-FROM-PDEs) project, by PRIN ``Reduced Order Models for Environmental and Urban flows'' (ROMEU), and by INdAM GNCS.

%% file: sections/appendix.tex
\section{Appendix}
\subsection{Machine learning of the temporal eddy viscosity coefficients}\label{appendix}

The main building blocks of the feed-forward NNs contain four layers:  an input layer which is the \textit{temporal vector coefficients of the velocity}, two hidden layers of 10 units each followed by an ELU (exponential linear unit) activation function,  and the output layer which is the \textit{temporal vector coefficients of the eddy viscosity.} 
The first 50\% of the data in the time series is used for training (30\%) and validation (20\%) and the remaining for testing. The model is trained over 500 epochs. 

For LSTM-RNNs, an LSTM Encoder-Decoder model is used for training,  validation, and test on the data set.  The architecture of such a LSTM-Encoder-Decoder model has one LSTM layer in the encoder part and another LSTM layer --- in addition to one hidden layer of 5 units --- in the decoder part. The reason for the choice is that this model has been used extensively for sequence-to-sequence prediction in the literature for NLP (natural language processing). The setup parameters of the LSTM are reported in \cref{tab:rnnsparameters}.
\begin{table}[H]
\centering
\begin{tabular}{|p{1.4cm}|p{1cm}|p{1cm}|p{1.25cm}|p{1cm}|p{1cm}|p{1cm}|p{1cm}|p{1cm}|}
\hline
 Parameter & Hidden  size &  learning rate &   optimizer & LSTM layers & Batch size & Epochs & input dim & output dim \\
\hline
Value & 1 & 1e-4 & ADAM & 1  & 5  & 500 & 5 & 5\\
\hline
\end{tabular}
\caption{LSTM hyper-parameters}
\label{tab:rnnsparameters}
\end{table}

\begin{figure}[H]
\centering
\includegraphics[width=\linewidth]{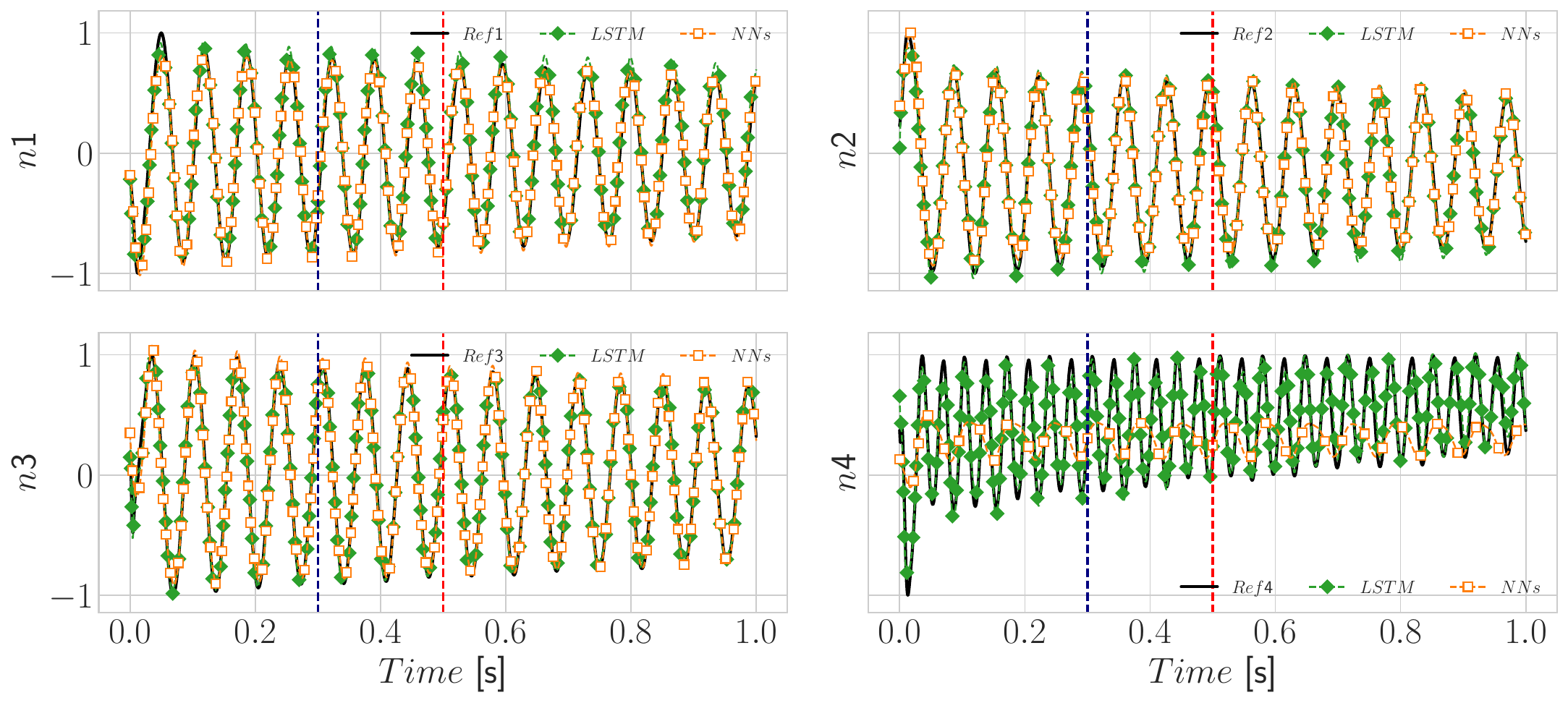}\\
\caption{Training, validation, and test of the first four POD temporal  coefficients of the eddy viscosity at the offline level using LSTM and NNs. The first half of the datasets (scaled between -1 and 1) is divided in two parts (30\% for training and 20\% for validation ) for training and validation on  both models. The remaining half is used for testing on both architectures}
\label{fig:lstm-ffnns}
\end{figure}